\documentclass[%
 reprint,
 amsmath,amssymb,
 aps,
]{revtex4-1}

\usepackage{graphicx}
\usepackage{dcolumn}
\usepackage{bm}
\newcommand{\ket}[1]{\vert#1\rangle}
\newcommand{\bra}[1]{\langle #1\vert}
\newcommand{\sand}[3]{\left\langle#1\vert#2\vert#3\right\rangle}

\newcommand{\Bo}[1]{\noindent{\bf #1}}

\begin{document}

\preprint{AIP/123-QED}

\title[]{Quantum wave--particle superposition in a delayed-choice experiment}

\author{Kai Wang}
\author{Qian Xu}%
\author{Shining Zhu}
\author{Xiao-song Ma}
\homepage{E-mail:  Xiaosong.Ma@nju.edu.cn}
\affiliation{ 
	National Laboratory of Solid-state Microstructures \& School of Physics, Nanjing University 
}%
\affiliation{%
	Collaborative Innovations Center of Advanced Microstructures
}%

\date{\today}

\begin{abstract}	
Wave--particle duality epitomizes the counterintuitive character of quantum physics. A striking illustration is the quantum delayed-choice experiment, which is based on Wheeler's classic delayed-choice gedanken experiment, but with the addition of a quantum-controlled device enabling wave-to-particle transitions. Here, we realize a quantum delayed-choice experiment in which we control the wave and the particle states of photons and particularly the phase between them, thus directly establishing the created quantum nature of the wave--particle. We generate three-photon entangled states and inject one photon into a Mach--Zehnder interferometer embedded in a 186-m-long two-photon Hong--Ou--Mandel interferometer. The third photon is sent 141~m away from the interferometers and remotely prepares a two-photon quantum gate according to independent active choices under Einstein locality conditions. We realize transitions between wave and particle states in both classical and quantum scenarios, and therefore tests of the complementarity principle that go fundamentally beyond earlier implementations.
\end{abstract}

\maketitle

\section*{\label{sec:level1}Introduction}
Quantum mechanics predicts effects that are in conflict with our everyday perceptions of nature. This is no better illustrated than in the wave--particle duality, where a single quantum system can simultaneously exhibit distinct and mutually exclusive features of a particle and a wave, respectively \cite{BOHR1928}. Hidden-variable theory was proposed to reconcile this conflict between quantum and classical physics. To rule out certain hidden variable theory, Wheeler proposed the so-called delayed-choice gedanken experiment to exclude the possibility that input photons can `know' the settings of the experiment and behave accordingly \cite{WHEELER19789,wheeler1984quantum}. The basic concept of Wheeler's delayed-choice (WDC) experiment is shown in Fig.1\textbf{a}. Single photons are injected into a Mach--Zehnder interferometer (MZI) consisting of two beam splitters (BS1 and BS2), a phase shifter ($\varphi$) and mirrors. The choice of whether or not to insert BS2 into the MZI is made by an external observer, a quantum random number generator (QRNG) for instance, and is delayed until after the input single photons pass through BS1 and hence enter into the MZI. The choice of inserting BS2 results in phase-dependent photon counts at the single-photon detectors (D1 and D2), which are the manifestations of the wave properties of the input single photons entering the interferometer. In contrast, by removing BS2 the counts at D1 and D2 are independent of the phase $\varphi$ and display anti-bunching behavior when the coincidence counts between D1 and D2 are measured. This leads to a seemingly paradoxical situation: Whether the input photons behave as a particle or a wave depends on the observer's delayed choice. 

Wheeler's gedanken experiment has been realized with photons~\cite{alley1986results,PhysRevA.35.2532,Baldzuhn1989,jacques2007experimental} and atoms~\cite{Manning2015} (for recent reviews, see refs~\cite{Shadbolt2014,RevModPhys.88.015005}).
As the configuration of the WDC experiment is controlled by the QRNG bit value (0 or 1), one can only probe the wave, the particle or the classical mixture of these two properties. The complementary interplay between particle and wave is well captured by the complementarity inequality~\cite{PhysLettA.128.391,PhysRevA.51.54,PhysRevLett.77.2154}. In 2011, Ionicioiu and Terno proposed a quantum version of delayed-choice experiment (QDC)~\cite{PhysRevLett.107.230406}, in which BS2 is substituted by a quantum-controlled beam splitter that can be in a superposition of being present and absent, as shown in Fig.1\textbf{b}. The configuration of the interferometer through which the system photon (S) propagates is determined by the quantum state of the control photon (C) by interactions. One advantage of the QDC experiment is that the experimentalist can arbitrarily choose the temporal order of the measurements on photons S and C. The QDC proposal and experimental realizations thereof are closely related to the well-known quantum eraser concept proposed by Scully and Dr\"uhl~\cite{PhysRevA.25.2208}.  Note that the coherence in a quantum-controlled beam splitter has also been theoretically analyzed using entropic uncertainty relation in a non-delayed-choice configuration~\cite{coles2014equivalence}. Moreover, a recent experiment established an alternative route to continuously morphing between the wave and particle behavior of photons, demonstrating that the superposition of the particle and wave states can be controlled for each of two photons, together with the degree of their wave-particle entanglement\cite{rab2017entanglement}.

The QDC concept invited broad experimental efforts across different systems~\cite{peruzzo2012quantum,kaiser2012entanglement,tang2012realization,PhysRevA.85.022109,PhysRevA.85.032121,PhysRevA.92.022126,PhysRevLett.115.260403,Liue1603159}. Very recently, a device-independent casual modeling framework for delayed-choice experiments has been theoretically proposed~\cite{PhysRevLett.120.190401}, within which conflicts between classical and quantum physics are certified through a dimension witness under assumptions. This proposal has subsequently been implemented in a series of independent experiments~\cite{2018arXiv180600211P,2018arXiv180600156H,2018arXiv180603689Y}. In all QDC experiments, photons C and S have to interact directly in order to facilitate the operation of the quantum-controlled gate, which therefore hinders the realization of the required relativistic separations between relevant events. Later, a novel version of the non-local quantum delayed-choice experiment was proposed with multiple entangled photons~\cite{ionicioiu2014wave}. The authors found the incompatibility between hidden-variable theories that simultaneously satisfy the conditions of objectivity, determinism and local independence of hidden variables and quantum mechanics describing entangled system~\cite{ionicioiu2014wave}. Here we report an experimental realization of the non-local delayed-choice experiment, which goes beyond both the WDC~\cite{WHEELER19789,wheeler1984quantum}and the QDC proposals~\cite{PhysRevLett.107.230406,ionicioiu2014wave}, and all previous demonstrations by showing the genuine quantum wave-particle superpositions. Our work ultimately enables novel aspects of wave--particle duality to be explored, in the form of controllable quantum superpositions of the wave and particle states.

\begin{figure}
	\includegraphics[width=8cm]{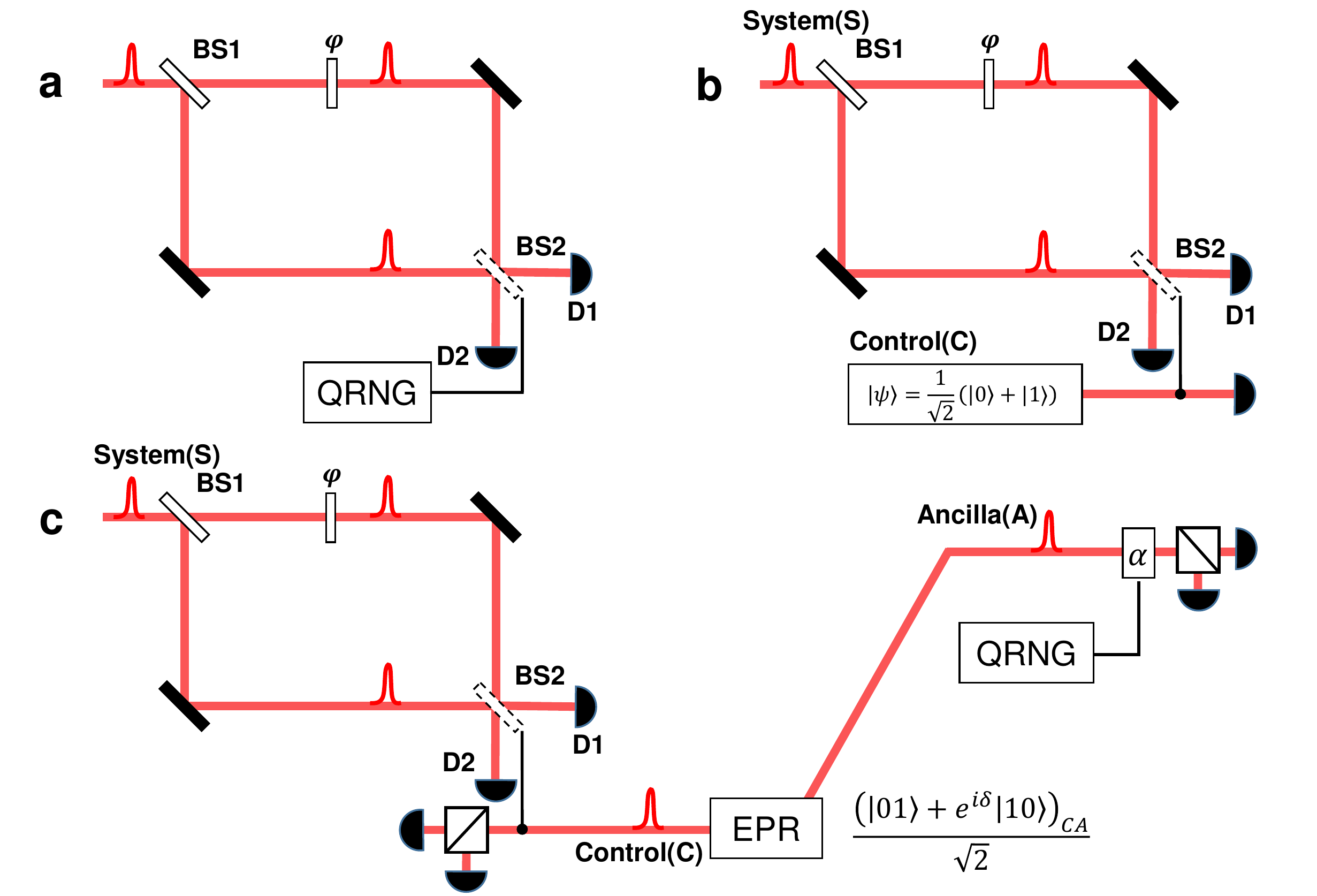}
	\caption{\textbf{ The evolution of delayed-choice experiments.} \noindent{\bf a}, Wheeler's delayed-choice experiment. The choice of inserting or removing the output beam splitter (BS2) is made by a quantum random number generator (QRNG) and delayed until after the single photon passed through BS1. \noindent{\bf b}, Quantum delayed-choice experiment. Compared to \noindent{\bf a}, BS2 is now replaced with a quantum beam splitter, the setting of which is determined by the superposition state of a control photon (C). \noindent{\bf c}, Non-local quantum delayed-choice experiment. The setting of BS2 is decided by the correlations of a control (C) and an ancilla photon (A), emitted from a Einstein-Podolsky-Rosen (EPR) entangled photon-pair source. The polarization of photon A is randomly projected onto the $\ket{H}$/$\ket{V}$ or the $\ket{\alpha}=\cos\alpha\ket{H}+\sin\alpha\ket{V}$/$\ket{\alpha^{\bot}}=\sin\alpha\ket{H}-\cos\alpha\ket{V}$ basis. This allows us to remotely prepare the quantum gate, BS2. }
\end{figure}

\section*{Scheme of the non-local quantum delayed-choice experiment}
The conceptual scheme of our experiment is shown in Fig.1\textbf{c}. Three single photons are involved: the system photon (S) that is sent through the interferometer and an entangled photon pair consisting of a control (C) and an ancilla (A) photon. Replacing the control photon with an entangled pair enables us to overcome the limitations of WDC and QDC experiments, as it allows us to simultaneously realize quantum control via remote quantum gate preparation and the relativistic separation between the relevant events. This is the novel aspect enabled by the multi-photon system we use in our experiment.

The procedure of our experiment is as follows. First, we prepare input states:
\begin{equation}
\ket{\psi^{i}_{SCA}}=\ket{V}_{S}\otimes\frac{(\ket{HV}+e^{i\delta}\ket{VH})_{CA}}{\sqrt{2}},
\end{equation}
where $\ket{\psi^{i}_{SCA}}$ is the initial state of photons S, C and A (as labelled in Fig.1\textbf{c}), $\ket{H}$ and $\ket{V}$ stand for the horizontally and vertically polarized states of single photons, respectively, and $\delta$ is the relative phase between the $\ket{HV}$ and $\ket{VH}$ terms in the entangled state of photons C and A. Second, we send photon S into the interferometer, in which the output beam splitter is a Hadamard gate controlled by photon C. However, unlike the original QDC proposal~\cite{PhysRevLett.107.230406}, where photon C is in a pure state ($\ket{H}$ or $\ket{V}$, or their arbitrary superpositions), here photons C and A are entangled and the individual quantum state of photon C is a mixed state. The configuration of the interferometer for photon S does not depend on the quantum state of photon C, but on the correlation of photons C and A. This means that we can remotely prepare a two-photon quantum gate operating on photons S and C by controlling and measuring photon A based on the final state of photons S,C and A: 
\begin{align}
\ket{\psi^{f}_{SCA}}=&\frac{1}{\sqrt{2}}\ket{\Bo{p}}_{S}\ket{H}_{C}\ket{V}_{A}+\frac{e^{i\delta}}{\sqrt{2}}\ket{\Bo{w}}_{S}\ket{V}_{C}\ket{H}_{A},
\end{align}
where $\ket{\Bo{p}}_{S}$ and $\ket{\Bo{w}}_{S}$ stand for the particle and wave states of photon S, respectively~\cite{PhysRevLett.107.230406,ionicioiu2014wave}. When photon S is in $\ket{\Bo{p}}_{S}=\frac{1}{\sqrt{2}}(\ket{H}-e^{i\varphi}\ket{V})$, it behaves as a particle and when it is in $\ket{\Bo{w}}_{S}=e^{i\varphi/2}(-i\sin\frac{\varphi}{2}\ket{H}+\cos\frac{\varphi}{2}\ket{V})$, it behaves as a wave. Note that the states $\ket{\Bo{p}}$ and $\ket{\Bo{w}}$ represent an operational way to describe the particle (no interference) and wave (interference) behavior of the photon, and these two states are in general not orthogonal. Such operational definitions allow us to conveniently show the continuous morphing between the different wave-particle dual nature. We can probe the wave property, the particle property, classcial mixture and quantum superpositions of these two properties by projecting photon A onto the basis $\ket{\alpha}=\cos\alpha\ket{H}+\sin\alpha\ket{V}$, $\ket{\alpha^{\bot}}=\sin\alpha\ket{H}-\cos\alpha\ket{V}$, $\alpha\in[0,\frac{\pi}{2}]$. 

As shown in Fig.1\textbf{c}, a QRNG generates a bit string of 0 or 1 to randomly choose a projection angle equal to $\alpha$ or $0$ for photon A. When the QRNG gives a bit value of 0, then the projection angle equals $\alpha$. For the sake of simplicity, from now on we only discuss the results of photon A being projected into $\ket{\alpha^{\bot}}_{A}$. The complementary results of photon A being projected into $\ket{\alpha}_{A}$ are presented in Supplementary Information. When photon A is projected into $\ket{\alpha^{\bot}}_{A}$, photons S and C are in the state $\ket{\psi^{f}_{SC}}=-\cos\alpha\ket{\Bo{p}}_{S}\ket{H}_{C}+e^{i\delta}\sin\alpha\ket{\Bo{w}}_{S}\ket{V}_{C}$, as derived from Eq.(2). The detailed derivations are provided in Methods. Taking the trace over photon C gives a classical mixed state of photon S: 
\begin{equation}
\rho_S=\cos^2\alpha\ket{\Bo{p}}\bra{\Bo{p}}+\sin^2\alpha\ket{\Bo{w}}\bra{\Bo{w}}.
\end{equation}
For this classical case, the probability of obtaining photon S in $\ket{H}$ conditionally on projecting photon A into $\ket{\alpha^{\bot}}$, $P_{C}(\varphi,\alpha)=P_{S=\ket{H}|A=\ket{\alpha^{\bot}}}(\varphi,\alpha)$, is a function of $\varphi$ and $\alpha$: 
\begin{equation}
P_{C}(\varphi,\alpha,\delta)=\frac{1}{2}\cos^{2}\alpha+\sin^{2}\alpha\sin^{2}\frac{\varphi}{2}~.
\end{equation}
Note that $P_{C}(\varphi,\alpha)$ does not depend on $\delta$, the phase between the wave and particle states, manifesting its classical nature. By scanning both $\varphi$ and $\alpha$, we obtain the distribution of $P_{C}(\varphi,\alpha)$. 

Based on Eq.(2), by projecting photon A on $\ket{\alpha^{\bot}}$ and C on $\ket{-}$, where $\ket{\pm}=(\ket{H}\pm\ket{V})/\sqrt{2}$ stand for the diagonally and anti-diagonally linear polarization states, respectively, we obtain the renormalized state of photon S:
\begin{equation}
\ket{\psi_S}=C_{S}(\cos\alpha\ket{\Bo{p}}+e^{i\delta}\sin\alpha\ket{\Bo{w}}),
\end{equation}
which is a quantum superposition of $\ket{\Bo{p}}$ and $\ket{\Bo{w}}$ representing the superposition of the `ability' and `inability' to produce interference. Note that the normalized coefficient is $C_{S}=(1-\sqrt{2}\cos\alpha\sin\alpha\cos\varphi\cos\delta)^{-1/2}$. The resulting probability of obtaining photon S in $\ket{H}$ conditionally on photon C in the state $\ket{-}$ and photon A being in the state$\ket{\alpha^{\bot}}$,\\ $P_{Q}(\varphi,\alpha,\delta)=P_{S=\ket{H}|C=\ket{-}, A=\ket{\alpha^{\bot}}}(\varphi,\alpha,\delta)$, is
\begin{align}
\nonumber
P_{Q}(\varphi,\alpha,\delta)=&|C_{S}|^{2}[\frac{1}{2}\cos^{2}\alpha+\sin^{2}\alpha\sin^{2}\frac{\varphi}{2}\\
&+\sqrt{2}\cos\alpha\sin\alpha\sin\frac{\varphi}{2}\sin(\delta+\frac{\varphi}{2})].
\end{align}
It is clear from Eq.(6) that $P_{Q}$ depends on the phase $\delta$, displaying its quantum nature.
\begin{figure*}
	\includegraphics[width=17cm]{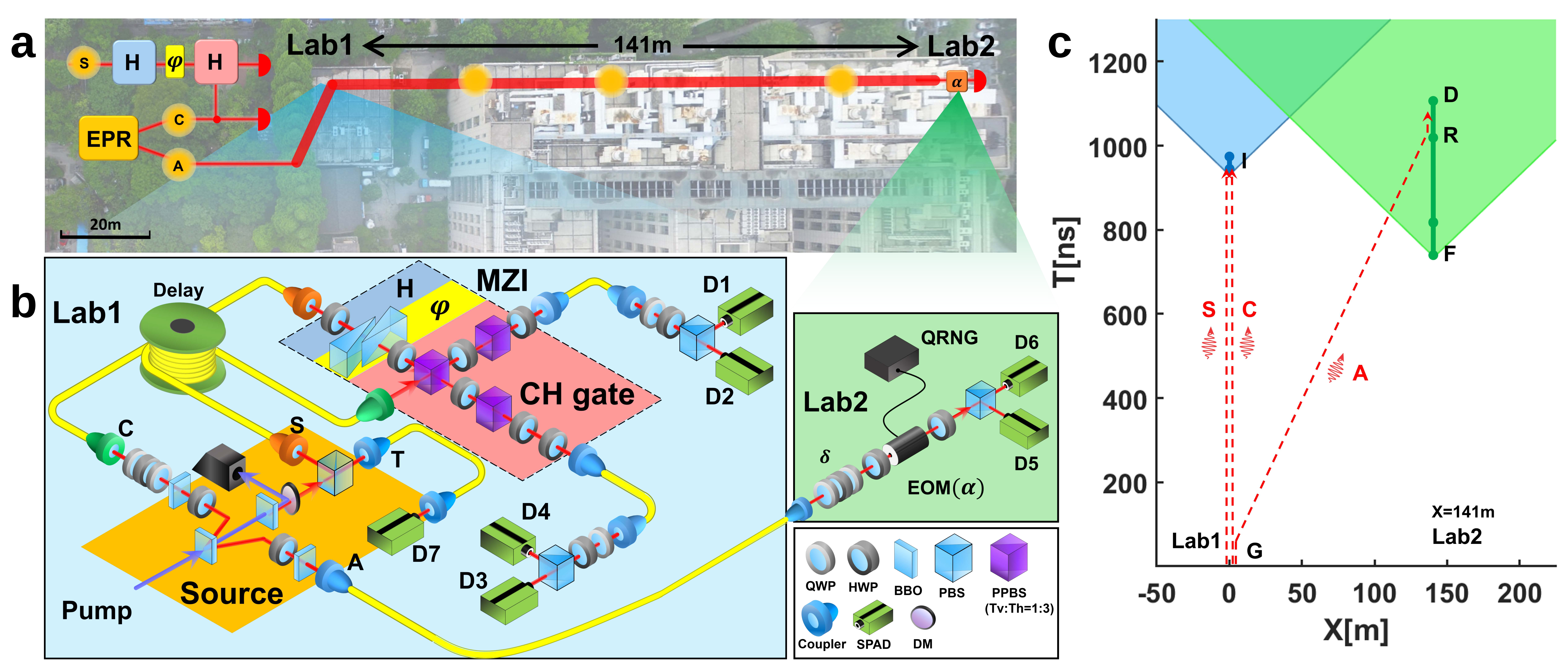}
	\caption{\label{fig:wide} \textbf{Experimental configuration.} \textbf{a}, Birds-eye view of the experiment. The experiment setup is distributed over two laboratories in the Tangzhongying building at Nanjing University, Lab1 and Lab2, which are separated by a line-of-sight distance of about 141~m. In Lab1, we generate the initial state $|\psi^{i}_{SCA}\rangle$. Photon S propagates through a MZI with a Hadamard gate (BS2 in Fig.1\textbf{c}) controlled by the entanglement of photons C and A generated by the source (EPR). Photon A is sent to Lab2 through a 215m-long optical fiber and its polarization  is measured there. \textbf{b}, Experimental setup. In Lab1, a source of polarization-entangled photon pairs generates photons A and C in BBO1 and a collinear photon-pair source generates photons S and T in BBO2. Photon S interferes with photon C on PPBS with perfect transmission Th for horizontal polarization and transmission Tv = 1/3 for vertical polarization in both output		modes after propagating through 186 m fibre delays, HWPs and a SBC($\varphi$). Photon A is sent to Lab2, where the random polarization rotation is implemented using an electro-optical modulator (EOM) controlled by a QRNG. All photons are detected by single-photon avalanche diodes (SPADs). See text for definitions of the components. \textbf{c}, Space-time diagram. We relativistically separate the interference events (I) of photon S with respect to the choice events and the polarization projection events (F, R, D) of photon A. }
\end{figure*}
\section*{Realization of a quantum delayed-choice experiment under Einstein's locality condition}
In Fig. 2, we show our experiment in detail. The experiment set-up is distributed over two laboratories (Lab1 and Lab2), which are connected with optical single-mode fiber cables and electrical coaxial cables. In Lab1, femtosecond pulses (central wavelength 808~nm) from a Ti:sapphire laser are up-converted to blue pulses (central wavelength 404~nm). These blue pulses are used to generate four photons (photons S, C, A as defined in Fig.1\textbf{c} and an additional trigger photon T) via type-II spontaneous parametric down-conversion in two $\beta$-barium borate (BBO) crystals placed in sequence~\cite{zukowski1995entangling}. Photons C and A are generated in BBO1, which is in the non-collinear configuration, and are in entangled states with tunable phase $\delta$ \cite{PhysRevLett.75.4337}. Photons S and T are generated in BBO2, which is in the collinear configuration, and are in product states $|VH\rangle_{ST}$. The detection of photon T projects photon S into a single-photon state. We couple all four photons into single-mode fibers for later manipulation and detection. Photon S is then sent through a 186 m single-mode fiber coil and is guided to the polarization MZI. The polarization MZI consists of one half-wave plate (HWP) orientated at an angle of $22.5^{\circ}$, a Soleil--Babinet Compensator (SBC) and a quantum-controlled Hadamard (CH) gate. The SBC introduces a relative phase $\varphi$ between the $|H\rangle$ and $|V\rangle$ states. The CH gate consists of two HWPs with their fast axes oriented at an angle of $11.25^{\circ}$ and a controlled Pauli-Z (CZ) gate between them. The CZ gate is realized through three partially polarizing splitters (PPBS) and four HWPs~\cite{PhysRevLett.95.210504,PhysRevLett.95.210505,PhysRevLett.95.210506}. To achieve a successful quantum CH gate, photons S and C have to arrive at the first PPBS simultaneously and interfere. Therefore, photon C is also passing through a 186 m single-mode fiber. For the details on the 186 m-fiber Hong-Ou-Mandel interferometer~\cite{PhysRevLett.59.2044}, see Supplementary Information. 

The key advance in our experiment, compared to previous QDC demonstrations, is that we realize a QDC experiment with multiple entangled photons and under Einstein's locality condition. To achieve this, we separate in space the events of random choice deciding whether to project photon A into the $|\alpha\rangle/|\alpha^{\bot}\rangle$ or the $|H\rangle/|V\rangle$ basis
and the events marking photon S entering into, propagating through and exiting from the MZI. Such separation is obtained by guiding photon A through a single-mode fiber from Lab1 to Lab2, which are separated by a line-of-sight distance of about 141~m. In Lab 2,  we use an EOM controlled by a QRNG, which generates random bits to set, with approximately equal probability, the polarization-analysis basis for photon A to be $|\alpha\rangle/|\alpha^{\bot}\rangle$ or $|H\rangle/|V\rangle$. All photons are filtered with interference filters (IF, 3-nm bandwidth). Seven single-photon avalanche detectors detect four photons in different spatial modes. 

The space-time diagram of our experiment is shown in Fig. 2\textbf{c}. At the origin, we generate four photons in Lab1 (event \Bo{G}). Photons S and C are delayed in two 186-m fibers (930 ns delay in time) in Lab1 before entering the MZI. After that, photon S interferes in the MZI (event \Bo{I}) controlled by photon C. In addition, a delay of approximately 29~ns is introduced for photons S and C by short fibers and free-space optics. Photon A is transmitted through a 215-m-long fiber (1075-ns delay in time) to Lab2. The QRNG has a repetition rate of 5~MHz, which means that the random bit and hence the settings of the EOM can change every 200~ns. The moment QRNG starts to generate a random bit is defined as event F (upon receiving a trigger signal). The electric signal of the QRNG is shaped with a discriminator and sent to the EOM (event \Bo{R}), which modulates the polarization state of photon A. We then analyze and detect the polarization of photon A (event \Bo{D}). The entire process between event \Bo{R} to \Bo{D} takes 88~ns. See Supplementary information for details regarding timing. The events relating to the interference of photon S in the MZI are space-like separated from the choice events and the polarization-projection events affecting photon A. In Lab1, we measure various three-photon coincidence counts between photons S, C and T. In Lab2, we perform polarization measurements on photon A and categorize the detection results according to the clicked detectors and bit value of the QRNG. Then after the measurement results have been irreversibly recorded with single-photon detectors in the two respective labs, we compare the results of S-C-T with A to obtain the conditional probability of photon S. These space-time arrangements ensure our realization of a QDC experiment under strict Einstein locality.

\section*{Characterization of quantum wave-particle superpositions}
\begin{figure*}
	\includegraphics[width=17cm]{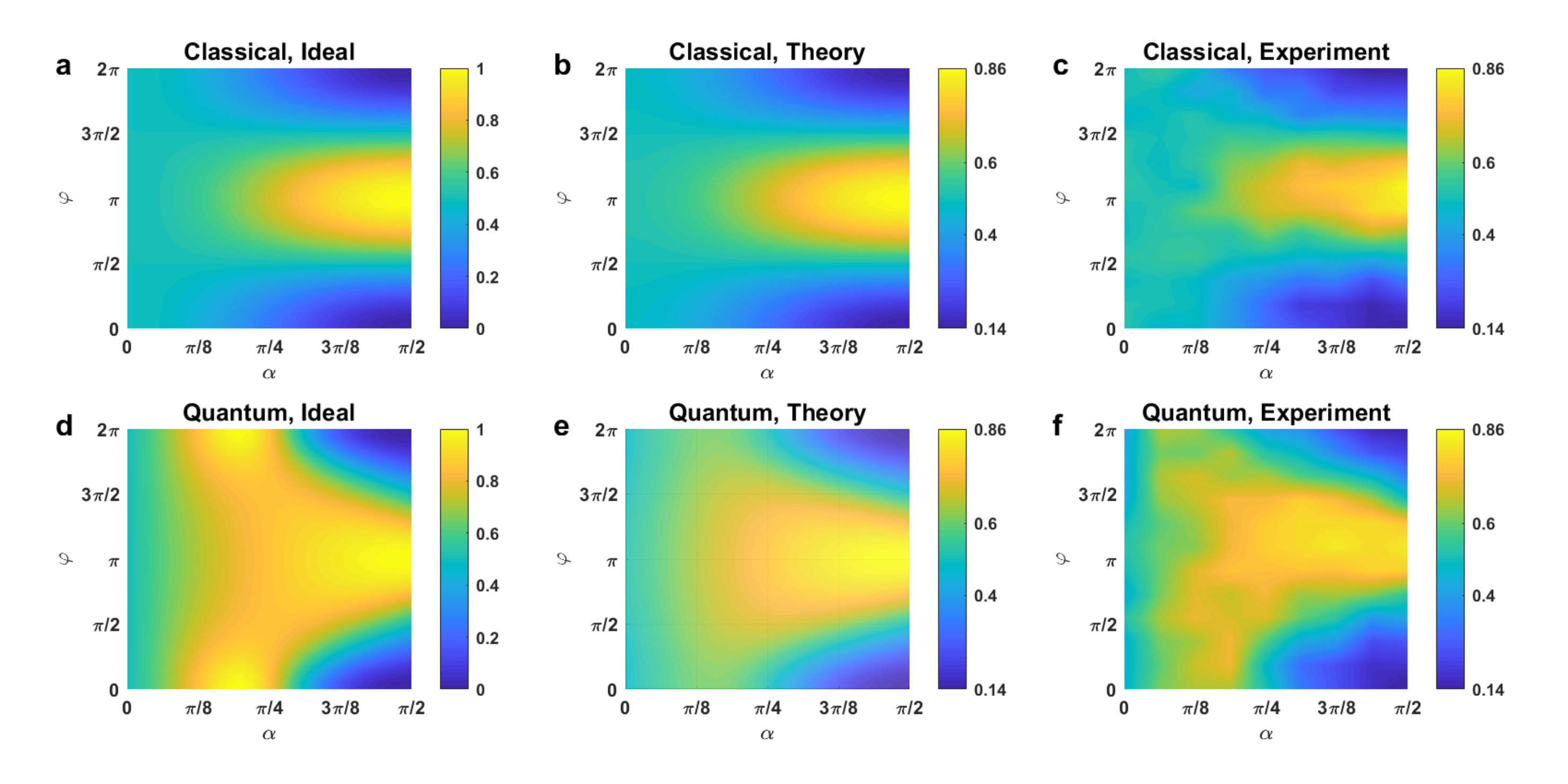}
	\caption{\label{fig:wide}\textbf{Continuous transitions between particle and wave states in both classical and quantum scenarios.} \textbf{a-c}, The simulated ideal, theoretically expected and experimentally measured probabilities $P_{C}(\varphi,\alpha)$ for a classical mixture of particle and wave states. They are obtained by scanning the phase $\varphi$ of MZI for photon S and polarization rotation angle $\alpha$ of photon A. \textbf{d-f}, The simulated ideal, theoretically expected and experimentally measured probabilities $P_{Q}(\varphi,\alpha,\delta)$ for a quantum coherent superposition of particle and wave states with $\delta=0$. In \textbf{c} and \textbf{f}, we have experimentally performed the 2D scan with nine values of $\alpha$ and twelve values of $\varphi$, which are equally distributed from 0 to $\pi/2$ and 0 to 2$\pi$, respectively. The intermediate values between each step are linearly interpolated and plotted accordingly. The error bars in \textbf{c} and \textbf{f} are derived from Poissonian statistics and range from 0.013 to 0.046 and 0.02 to 0.063, respectively.}
\end{figure*}

Photon S can be a particle or a wave, or a classical mixture of the two if we perform the corresponding measurements on photons C and A, as shown in Eq.(3). In Fig. 3\textbf{a}, we plot the ideal conditional probability of photon S exiting the polarization MZI in horizontal polarization as a function of the phase $\varphi$ of the MZI and the polarization projection angle $\alpha$ for photon A, as shown in Eq. 4. Here we assume that the correlation of the photon source and the visibility of interference are unity. At $\alpha=0$ or $\pi/2$, photon S behaves as a particle or a wave, respectively, and hence no or full interference is seen. For the angles between $0$ and $\pi/2$, photon S is in a classical mixed state of particle and wave. In this regime, the complementary principle is valid and can be quantitatively characterized by the complementarity inequality, as has been shown experimentally in the context of delayed-choice experiments~\cite{PhysRevLett.100.220402,kaiser2012entanglement,Ma1221}. In Fig. 3\textbf{b}, \textbf{c}, we show the expected and experimentally measured results, which agree well with each other. Note that the parameters used to calculate the probability distributions shown in Fig. 3\textbf{b} are based on the values obtained from independent experimental measurements. Comparing to Fig. 3\textbf{a}, there is noticeable visibility degradations, which is mainly due to experimental imperfections as quantitatively explained in Supplementary Information.

More intriguingly, photon S can be in a quantum superposition of its particle and wave states. As shown in Eq.(5), we can probe this superposition by projecting photon C and A onto the $\ket{+}/\ket{-}$ and $\ket{\alpha}/\ket{\alpha^{\bot}}$ bases, respectively. In Fig. 3\textbf{d-f}, we show the ideal, the theoretically expected and the experimental results of the conditional probability of photon S exiting the polarization MZI in horizontal polarization as a function of $\varphi$ and $\alpha$ (Eq.(6) with $\delta=0$), respectively. At $\alpha=0$ or $\pi/2$, photon S behaves as a particle or a wave, which is the same as in the classical case. However, distinct and subtle interference effects appear between $0$ and $\pi/2$, which are the signatures of genuine quantum superpositions. For detailed analysis and comparisons between the classical and quantum cases, see Supplementary Information.
\begin{figure*}
	\includegraphics[width=17cm]{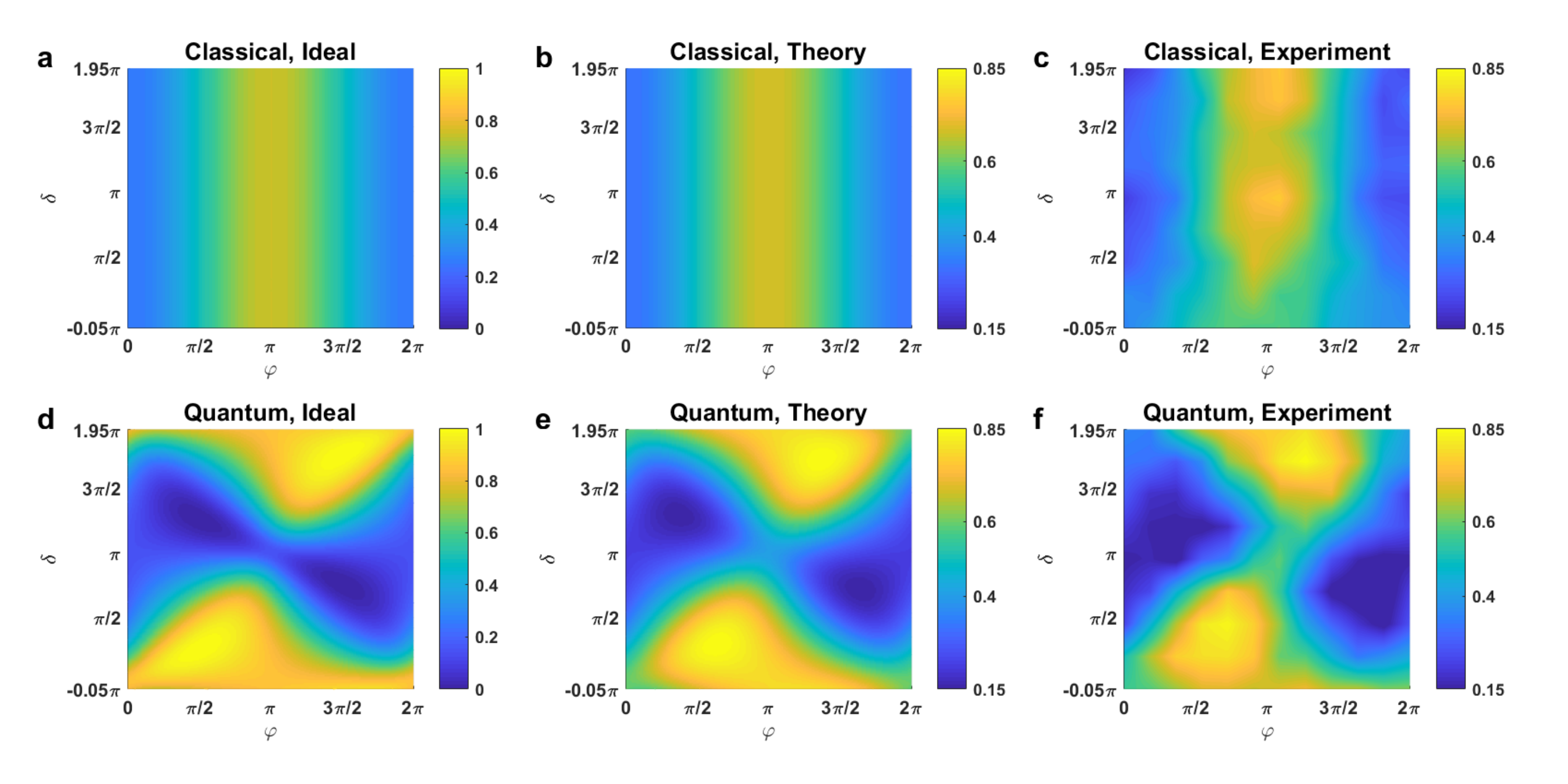}
	\caption{\label{fig:wide}\textbf{Witnessing the wave--particle quantum superpositions.} We fix $\alpha$ at $\pi/4$ and measure $P_{C}(\varphi,\delta)$ and $P_{Q}(\varphi,\delta)$ as functions of $\varphi$ and $\delta$.  \textbf{a-c}, The simulated ideal, theoretically expected and experimentally measured probabilities $P_{C}(\varphi,\delta)$ for a classical mixture of particle and wave states. \textbf{d-f}, The simulated ideal, theoretically expected and experimentally measured probabilities $P_{Q}(\varphi,\delta)$ for a quantum  superposition of particle and wave states. $P_{C}(\varphi,\delta)$ is independent of $\delta$, whereas $P_{Q}(\varphi,\delta)$ is strongly dependent on $\delta$, manifesting the quantum nature of the superposition of wave-particle states. In \textbf{c} and \textbf{f}, we have performed the 2D scan with nine values of $\delta$ and twelve values of $\varphi$, which are equally distributed from -0.05$\pi$ to -1.95$\pi$ and 0 to 2$\pi$. The intermediate values between each step are linearly interpolated and plotted. The error bars in \textbf{c} and \textbf{f} are derived from Poissonian statistics and range from 0.02 to 0.043 and 0.02 to 0.074, respectively.}
\end{figure*}

The most direct proof for the quantum nature of the wave--particle superposition is to show that the result is sensitive to the relative phase between the wave and particle states, $\delta$ (see Eq. (6)). For classical mixtures, no such dependence exists (see Eq. (4)). In order to probe such phase dependence, we fix $\alpha$ at $\pi/4$, and measure $P_C$ and $P_Q$ as functions of $\varphi$ and $\delta$. The results are shown in Fig. 4\textbf{a}-\textbf{c} for the classical mixture between the wave and the particle states, which are insensitive to the phase $\delta$. On the other hand, we show the results for genuine quantum coherent superpositions between wave and particle states in Fig. 4\textbf{d}-\textbf{f}. We stress that in the classical case $P_{C}$ remains sensitive to $\varphi$ (Fig. 4\textbf{c}), which means that the well-known complementarity inequality can be verified. However, verification of the complementarity inequality ~\cite{PhysLettA.128.391,PhysRevA.51.54,PhysRevLett.77.2154,PhysRevLett.100.220402,kaiser2012entanglement,Ma1221}  is not sufficient for proving that wave and particle states exist in a quantum superposition state. This is the key insight provided by the present work. 

In conclusion, we have realized a non-local quantum delayed-choice experiment with multiphoton entangled states. We have shown that a single photon can be a particle, a wave, a classical mixture of particle and wave as well as a quantum superposition of particle and wave. Its property depends on the choice of the correlation measurements of two other photons, even if that choice is made at a location and a time such that it is relativistically separated from the photon entering into, propagating through and exiting from the MZI. By doing so, we have a situation where no interaction between photon A, and photons S and C, not even a hypothetical one that would propagate with the speed of light, would allow photon A to determine the property of photon S. This is a strong constraint set by the theory of special relativity. Our work provides the realization of wave--particle quantum superposition and the first implementation of a full QDC experiment under strict Einstein locality conditions. 

\section*{Methods}
The choice-dependent final state of photons S,C,A can be written as $\ket{\psi^{f}_{SCA}}:$
\begin{align}
\nonumber\ket{\psi^{f}_{SCA}}=&\frac{1}{\sqrt{2}}[\sin \alpha\ket{\Bo{p}}_{S}\ket{H}_{C}+e^{i\delta}\cos \alpha\ket{\Bo{w}}_{S}\ket{V}_{C}]\ket{\alpha}_{A}\\
+&\frac{1}{\sqrt{2}}[-\cos\alpha\ket{\Bo{p}}_{S}\ket{H}_{C}+e^{i\delta}\sin\alpha\ket{\Bo{w}}_{S}\ket{V}_{C}]\ket{\alpha^{\bot}}_{A}~,
\end{align}
where we project photon C onto the $\ket{H}/\ket{V}$ basis. On the other hand, if we project photon C onto the $\ket{+}/\ket{-}$ basis. The final state is in the form of:
\begin{align}
\nonumber \ket{\psi^{f}_{SCA}}=&\frac{1}{2}[(\sin\alpha\ket{\Bo{p}}_{S}+e^{i\delta}\cos\alpha\ket{\Bo{w}}_{S})\ket{+}_{C}\\
\nonumber
&+(\sin\alpha\ket{\Bo{p}}_{S}-e^{i\delta}\cos\alpha\ket{\Bo{w}}_{S})\ket{-}_{C}]\ket{\alpha}_{A}\\
\nonumber
+&\frac{1}{2}[(-\cos\alpha\ket{\Bo{p}}_{S}+e^{i\delta}\sin\alpha\ket{\Bo{w}}_{S})\ket{+}_{C}
\\&-(\cos\alpha\ket{\Bo{p}}_{S}+e^{i\delta}\sin\alpha\ket{\Bo{w}}_{S})\ket{-}_{C}]\ket{\alpha^{\bot}}_{A}.
\end{align}

\section*{Acknowledgements}
The authors thank J. Kofler and \v{C}. Brukner for helpful discussions. This research is supported by the National Key Research and Development Program of China (2017YFA0303700), National Natural Science Foundation of China (11690032, 11674170 and 11621091), NSF Jiangsu province (No. BK20170010), the program for Innovative Talents and Entrepreneur in Jiangsu, and the Fundamental Research Funds for the Central Universities.

\appendix

\section{Lab1: Experimental details of the Mach-Zehnder and Hong-Ou-Mandel interferometers}

The goal of our experiment is to realize a Mach-Zehnder interferometer (MZI) with the quantum controlled Hadamard (CH) gate, acting as the output beam splitter (BS2 in Fig. 1\textbf{C} of the main text). In order to realize this goal, we embed the MZI into a two-photon Hong-Ou-Mandel interferometer (HOMI), which enables us to realize the CH gate based on measurement-induced nonlinearity~\cite{PhysRevLett.95.210504,PhysRevLett.95.210505,PhysRevLett.95.210506}. The schematic of our setup in Lab1 is shown in Fig. 5. 
\begin{figure}[h]
	\includegraphics[width=8cm]{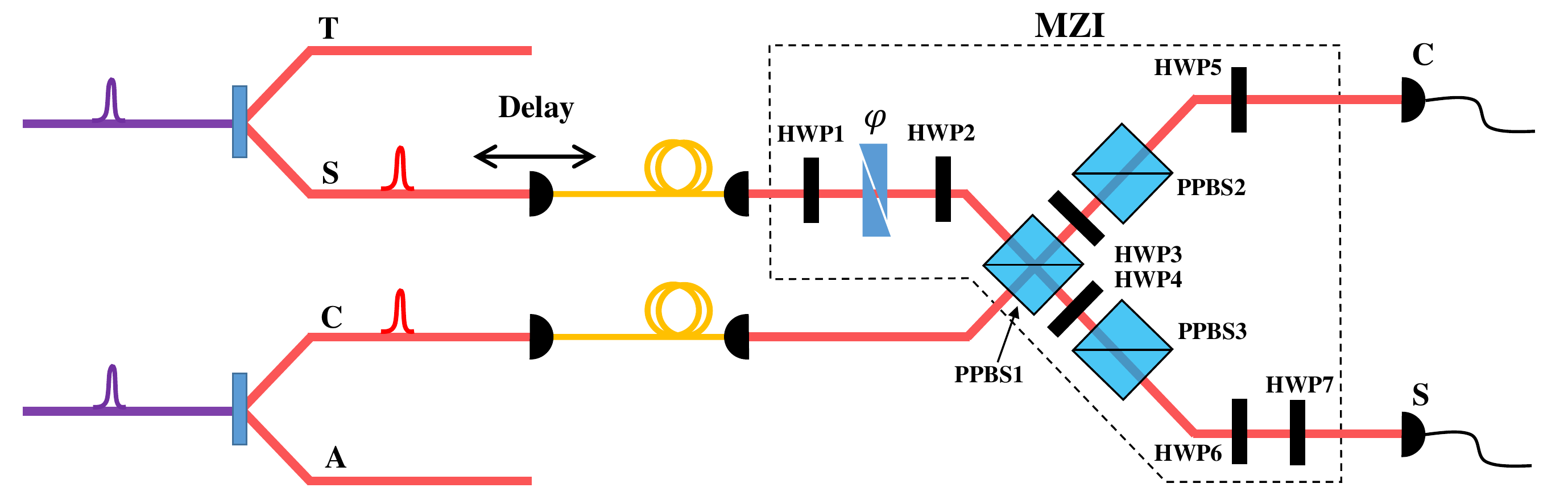}
	\label{Fig.5}
	\caption{Simplified experimental setup in Lab1. The Mach-Zehnder Interferometer (MZI) for photon S is realized with several polarization optical elements, as stated in the text. We introduce delay for photon S with respect to photon C using a motorized translation stage which moves along the optical path to scan HOM interference pattern.}
\end{figure}

We use a polarization Mach-Zehnder Interferometer (MZI), in which the first beam splitter (BS1 in Fig. 1\textbf{C} of the main text) is realized with a half-wave plate (HWP1) with its optical axis orienting along $22.5^{\circ}$. Phase $\varphi$ is adjusted by Soleil-Babinet compensator (SBC). A controlled-Pauli-Z (CZ) gate is made by three partial polarization beam splitters (PPBS1-3) and four half-wave plates (HWP3-6). Two W gates, realized with HWP2 and HWP7 oriented along $11.25^{\circ}$, and the CZ gate constitute a controlled-Hardmard (CH) gate in the following form:
\begin{equation}
\Bo{CH_{CS}}=(\Bo{I}_{C}\otimes\Bo{W}_{S})\Bo{CZ}_{CS}(\Bo{I}_{C}\otimes\Bo{W}_{S}),
\end{equation}
which acts as $BS_2$ in Fig. 1\textbf{C} of the main text, where \\
\begin{align}
&\rm{CH\ gate}:
\Bo{CH_{CS}}=\begin{bmatrix}
& 1 & 0 & 0 & 0\\
& 0 & 1 & 0 & 0\\
& 0 & 0 & \frac{1}{\sqrt{2}} & \frac{1}{\sqrt{2}}\\
& 0 & 0 & \frac{1}{\sqrt{2}} & -\frac{1}{\sqrt{2}}\\
\end{bmatrix}\\
&\rm{W\ gate}:  
\Bo{W}_{S}=
\begin{bmatrix}
&\cos\pi/8 &\sin\pi/8\\
&\sin\pi/8 &-\cos\pi/8\\
\end{bmatrix}\\
&\rm{Controled-Z\ gate}:  
\Bo{CZ}_{CS}=\begin{bmatrix}
& 1 & 0 & 0 & 0\\
& 0 & 1 & 0 & 0\\
& 0 & 0 & 1 & 0\\
& 0 & 0 & 0 & -1\\
\end{bmatrix}\\
&\rm{Identity\ gate}:
\Bo{I}_{C}=
\begin{bmatrix}
&1 & 0 \\
&0 & 1 \\
\end{bmatrix}.
\end{align}
The subscripts represent photons being manipulated.

All the PPBSs we used in the CZ gate have the similar optical specifications with the ratio of horizontal and vertical polarized light's transmission coefficients equalling to $T_{H}:T_{V}=3:1$. To facilitate the CZ gate, we have to coherently overlap photons S and C on CZ. This is confirmed by the observations of HOM interference dip, which is realized by scanning the delay of photon S with respect to photon C. SBC introduces various birefringent phases, $\varphi$, for photon S as we adjust the thickness of SBC. We measure HOM interference patterns with phase $\varphi$ changing from $0$ to $2\pi$, shown in Fig.6 (\Bo{A-I}). These results show a relation between the position of HOM dip and phase $\varphi$, as in Fig.6 (\Bo{J}). According to this relation, we set different values of the delay (corresponding to the position of HOM dip) for different phases in our experiment. 
\begin{figure*}
	\includegraphics[width=17cm]{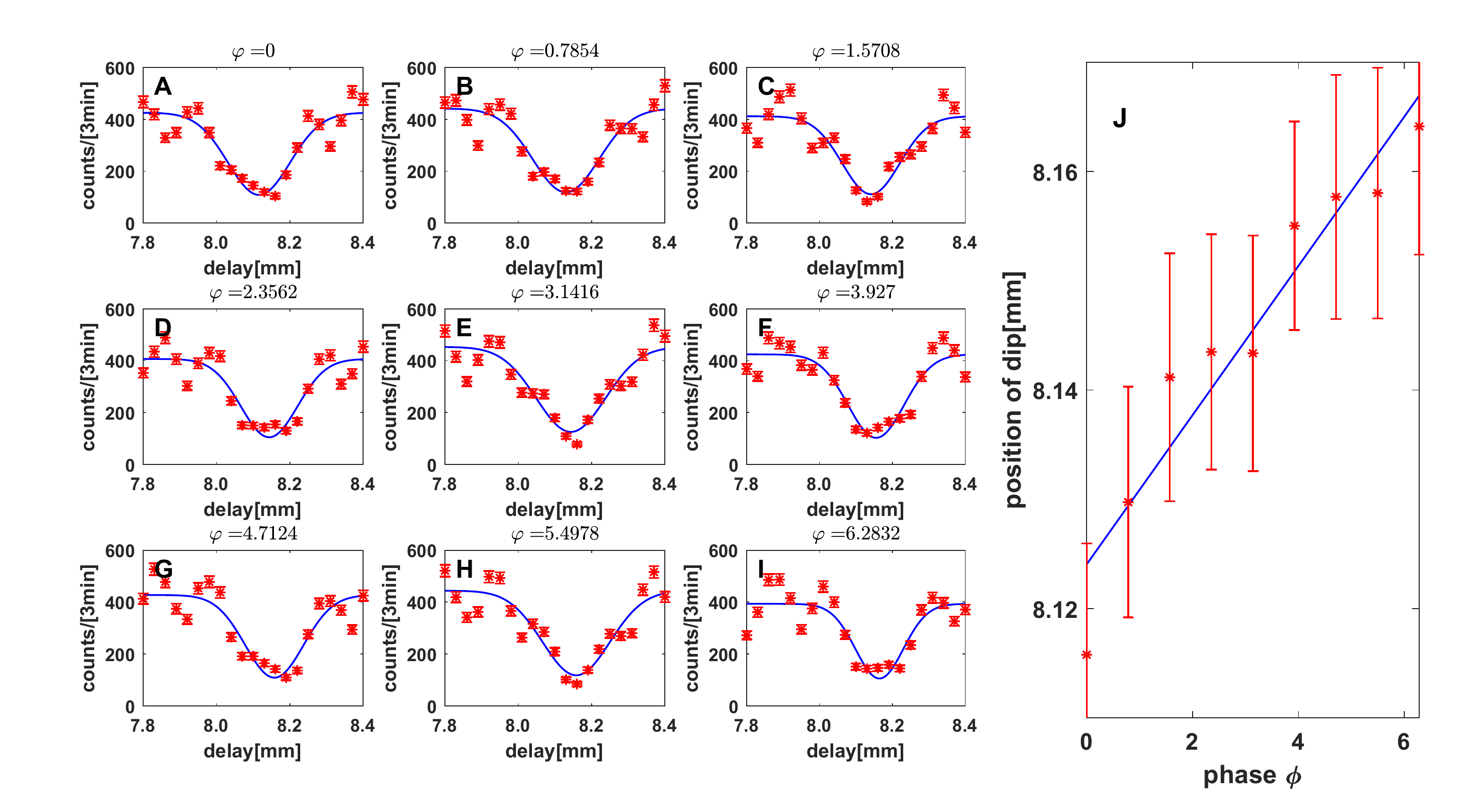}
	\caption{The relations between phase $\varphi$ and HOM-dip positions. (\Bo{A-I}) The HOM interference patterns have been measured at the different phase $\varphi$ from about $0$ to $2\pi$. Horizontal axis represents the position of the translation stage, which introduces the optical delay. Vertical axis represents the four-fold coincidence count $\ket{HVVH}_{TSCA}$. In ideal condition, HOM interference on PPBS gives a contrast value of 0.8 (C=(Max-Min)/Max). Fitting curves of our experimental data give the average contrast $0.738\pm0.084$. (\Bo{J}) Linear fit of the position of HOM dip as a function of phase $\varphi$. The error bar represent standard deviation and is obtained from curve fitting. Each point corresponds to one figure in (\Bo{A-I}).}
\end{figure*}

\section{Lab2: Experimental details of photon A's measurement setup}

\indent Our setup of Lab2 is shown in Fig.7. To implement fast optical switch between $\ket{H}/\ket{V}$ and $\ket{\alpha}/\ket{\alpha^{\bot}}$ basis, we use an electro-optic modulator (EOM) with optical axis oriented along $45^{\circ}$ relative to the laboratory coordinates and two HWPs at the angle $\theta$. When random number is 1, $\alpha=0^{\circ}$; when it is 0, $\alpha=4\theta-\frac{\pi}{2}$. After the rotation of angle $\alpha$, photon A is detected in $\ket{H}/\ket{V}$ basis. The details of electric and optical signal processing are shown in Fig.7 (\Bo{A}). A quantum random number generator (QRNG), driven by 5~MHz square-wave signals from function generator (FG), takes about 80~ns to response to the driving signal and then streams out random numbers as shown in Fig.7 (\Bo{B}). The random number signals (RNS) are divided into two identical copies, RNS1 and RNS2, respectively. RNS1 signals are shaped with a discriminator, amplified by the EOM driver and sent to EOM. EOM responses to the driving signal and modulates the transmitting light accordingly. Then we use two photon detectors (denoted as Detector H/V for horizontal/vertical polarization) on each output port of the PBS to analyze the polarization of output light. The whole process from quantum random number input to single-photon detection output takes 88~ns as shown in Fig.15 (\Bo{C}). RNS2 signals are used to make coincidence counts and hence we can categorize the detector signals: photons being modulated at the high or low level of random number signals, corresponding to \textit{I}/$\alpha$ gate operation. Then RNS2 and inverted RNS2 will both do \Bo{AND} operation with signal pulses from detectors and logic circuit counts the resulting four coincidence counts: Detector \Bo{H} with Identity(\textit{I}); Detector \Bo{H} with $\alpha$ gate; Detector \Bo{V} with Identity; Detector \Bo{V} with $\alpha$ gate. 

\begin{figure*}
	\includegraphics[width=15cm]{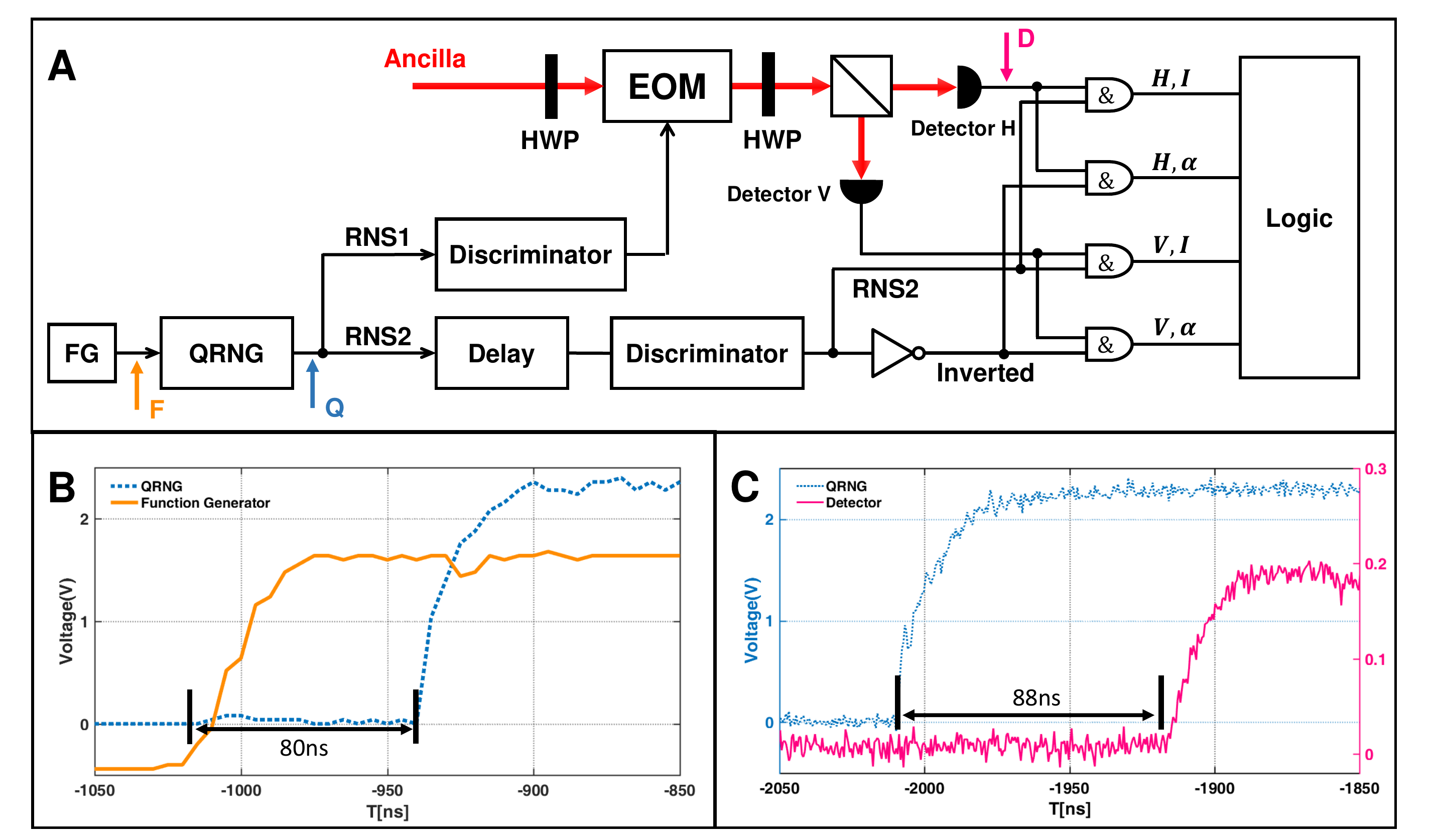}
	\caption{ Setup in Lab2. (\Bo{A}) Electric and optical signals processing scheme. A function generator (FG) sends signals to drive QRNG to generate random number number stream at the frequency of 5~MHz. The random number signals are divided into RNS1 and RNS2. RNS1 signals drive the EOM. RNS2 signals categorize the detector signals. Logic circuit analyses and counts the categorized detector signals. Three marked arrows F,Q,D represent the signals from FG,QRNG and Detector H being present at the corresponding locations. Their temporal separations are shown in Fig.7 (\Bo{B},\Bo{C}). (\Bo{B}) Response of QRNG to FG signal. Orange solid/blue dashed curve represents the trigger from the FG/the output random number signals from QRNG at the location of F/Q, respectively. QRNG takes about 80~ns to response the trigger signal. (\Bo{C}) Blue dashed and red solid curves respectively represent the random number signals from QRNG and the detection signals from Detector H, showing the modulation effect of EOM. The time interval of these two signals is about 88ns. The whole process from quantum random number input to photon detection output is contained in this period.}
\end{figure*}
In order to find the exact temporal delay of RNS2 signals for compensating the EOM and detector's response, we sent photons in $\ket{H}$ state into the EOM oriented at $45^{\circ}$ (Fig.8 (\Bo{A})). At the low level of random number signal, EOM will rotate the polarization states of photons to $\ket{V}$ state and Detector V fires; at the high level, the polarization states of photon keep unchanged and only Detector H fires as shown in Fig.8 (\Bo{C}). The signal pulses from Detector H will do \textbf{AND} operation with RNS2 and inverted RNS2 (Fig.8 \Bo{E},\Bo{F}) . When we find the right delay between RNS2 and detector pulses, there should be a maximum contrast between the output counts of the two \textbf{AND} gate. This is because that, at the high level of random number signals, the polarization states of the photon are mainly at $\ket{H}$. In our experiment, we obtained a maximum contrast about 18:1 with 5~MHz random number signal with the correct delay. This is shown with the single-photon counting results in Fig.9. Complementary illustrations of Detector \textbf{V} are shown in Fig.8. \Bo{G}, \Bo{H} and \Bo{I}.

\begin{figure*}
\includegraphics[width=15cm]{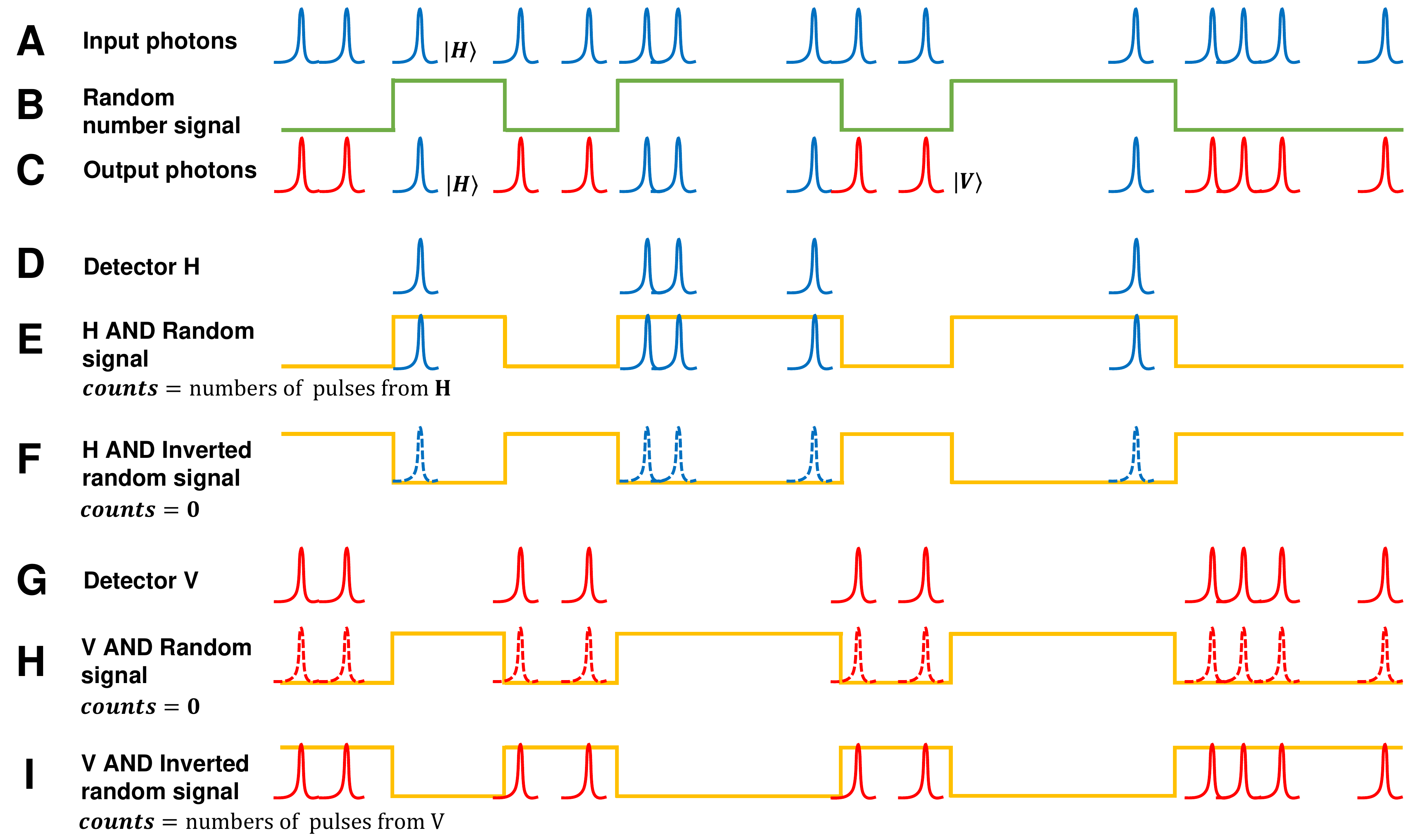}
	\caption{Schematic diagram of sorting single-photon detection signals with random number signals. (\Bo{A}) Horizontally polarized single photons, $\ket{H}$, are sent to EOM. (\Bo{B}) Random number signals from QRNG. (\Bo{C}) Polarization state of photons coming out from EOM. The polarization states are rotated to $\ket{V}$ (red) if the random bit is ``0'' and kept to $\ket{H}$ (blue) if the random bit is ``1''. (\Bo{D}) Signals from Detector H. (\Bo{E})/(\Bo{F}) The coincidence counts of signals from Detector H and RNS2/inverted RNS2 at the right delay. (\Bo{G}) Signals from Detector V. (\Bo{H})/(\Bo{I}) The coincidence counts of signals from Detector V and RNS2/inverted RNS2 at the right delay. }
\end{figure*}

\begin{figure}[h]
\includegraphics[width=8cm]{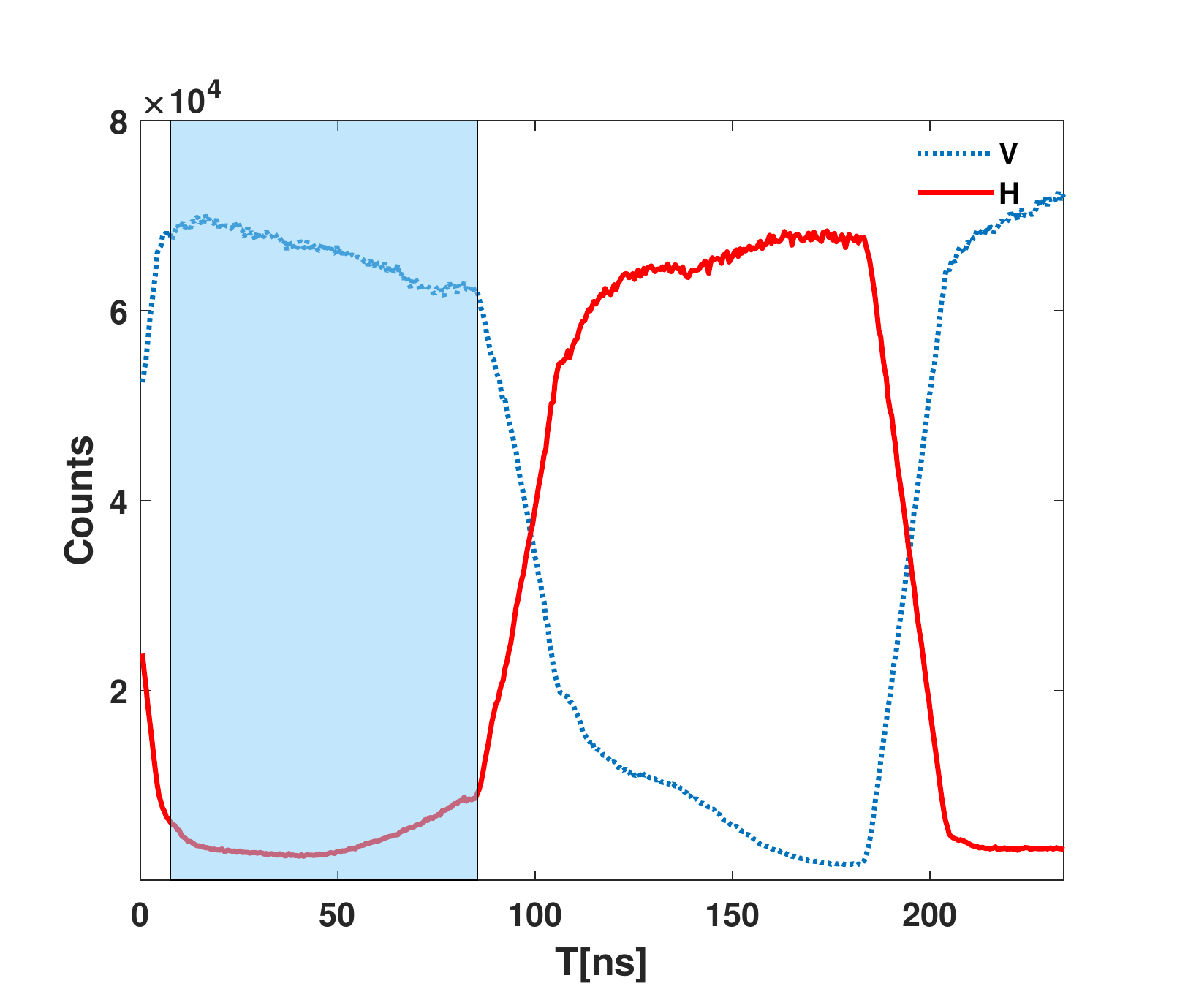}
\caption{Measurements of switching contrast with single photons. At the right delay, we send photons in horizontal polarization into EOM and measure the average contrast ratio using 5~MHz driving signal. \Bo{H} and \Bo{V} represent the photon counts of Detector H, V after the modulation of EOM. We take arithmetic mean of the contrast within the shading area, which is about $C=\frac{Count_{V}}{Count_{H}}=18:1$.}
\end{figure}

\section{Connection: Fibers and Cables}
Single-photon detector signals of D1-D4 and D7 in Lab1 are transmitted to Lab2 by five 210~m coaxial cables. The amplitude of pulses from detectors is 4.37V and decreases to 2.84V after passing the long cable while the rising/falling time keeps almost unchanged (6ns/10ns), respectively.
\begin{figure}[h]
	\includegraphics[width=8cm]{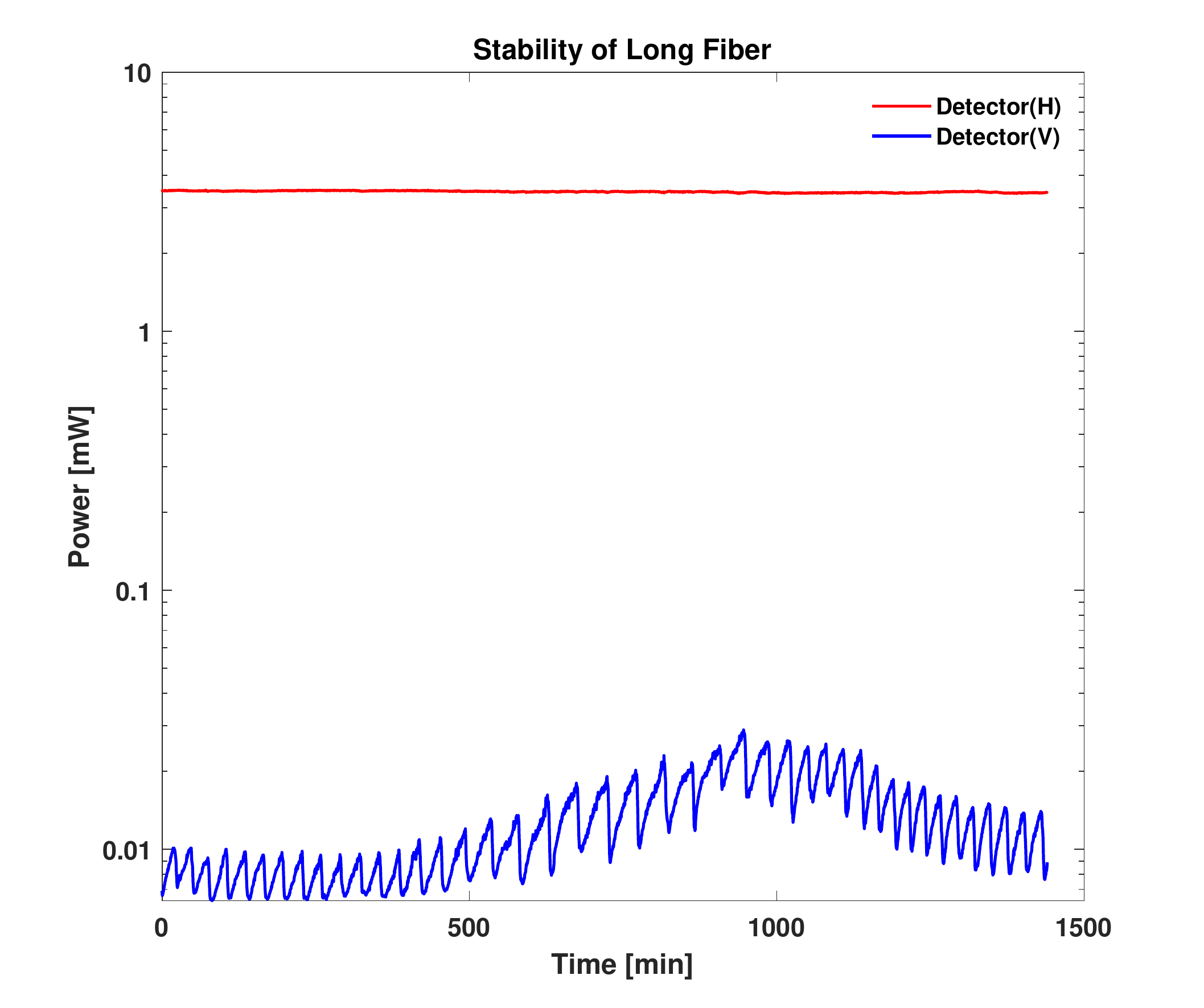}
	\caption{ We send horizontal-like polarized laser light into long fiber in Lab1 and analyze its polarization in Lab2 to monitor the polarization stability of the 215~m fiber across two labs. This measurement lasts for about 24h.}
\end{figure}

The long fiber used to transmit photon C is 780HP single-mode fiber. It's protected by Polyethylene pipe and a buffer tube. The polarization stability is measured with laser light and shown in Fig.10. Although there are small periodical modulations in power due to the inefficiency of the air conditioner temperature feed-back loop in Lab1, the polarization contrast is maintained to be more than 100 over 24 hours.

\section{Analysis and experimental results}
\subsection{Theoretical calculations and experimental data}
All the theoretical analyzations are based on the final state $\ket{\psi_{SCA}}$: 
\begin{align}
\nonumber \ket{\psi^{f}_{SCA}}=&\frac{1}{2}[(\sin\alpha\ket{\Bo{p}}_{S}+e^{i\delta}\cos\alpha\ket{\Bo{w}}_{S})\ket{+}_{C}\\ \nonumber
&+(\sin\alpha\ket{\Bo{p}}_{S}-e^{i\delta}\cos\alpha\ket{\Bo{w}}_{S})\ket{-}_{C}]\ket{\alpha}_{A}\\ \nonumber
+&\frac{1}{2}[(-\cos\alpha\ket{\Bo{p}}_{S}+e^{i\delta}\sin\alpha\ket{\Bo{w}}_{S})\ket{+}_{C}\\
&-(\cos\alpha\ket{\Bo{p}}_{S}+e^{i\delta}\sin\alpha\ket{\Bo{w}}_{S})\ket{-}_{C}]\ket{\alpha^{\bot}}_{A},
\end{align}
where we have projected photon A on detection basis $\ket{\alpha}/\ket{\alpha^{\bot}}$. Particularly, when QRNG gives a bit value of 1, $\alpha=0$ and  $\ket{\psi^{f}_{SCA}}=\frac{e^{i\delta}}{\sqrt{2}}\ket{\Bo{w}}_{S}\ket{V}_{C}\ket{H}_{A}-\frac{1}{\sqrt{2}}\ket{\Bo{p}}_{S}\ket{H}_{C}\ket{V}_{A}.$

Theoretical calculations and experimental results of $P_{C}$ and $P_{Q}$ for projecting photon S on $\ket{H}_{S}$ and $\ket{V}_{S}$ are shown in Fig.11 and Fig.12, respectively. Note that the data presented in Fig.11 (\textbf{B}) and (\textbf{D}) are identical to that shown in the Fig.3 (\textbf{C}) and (\textbf{F}) in the main text. The corresponding equations are:
\begin{align}
\nonumber&\Bo{Fig.11 (A)}\\
&P_C(\ket{H}_s)=P_{S=\ket{H}|A=\ket{\alpha^{\bot}}}(\varphi,\alpha)=\frac{1}{2}\cos^{2}\alpha+\sin^{2}\alpha\sin^{2}\frac{\varphi}{2}\\
\nonumber&\Bo{Fig.12 (A)}\\
&P_C(\ket{V}_s)=P_{S=\ket{V}|A=\ket{\alpha^{\bot}}}(\varphi,\alpha)=\frac{1}{2}\cos^{2}\alpha+\sin^{2}\alpha\cos^{2}\frac{\varphi}{2}\\
\nonumber&\Bo{Fig.11 (C)}\\
\nonumber&P_Q(\ket{H}_s)=P_{S=\ket{H}|C=\ket{-},A=\ket{\alpha^{\bot}}}(\varphi,\alpha,\delta)\\
&=\frac{\frac{1}{2}\cos^{2}\alpha+\sin^{2}\alpha\sin^{2}\frac{\varphi}{2}+\sqrt{2}\cos\alpha\sin\alpha\sin\frac{\varphi}{2}\sin(\delta+\frac{\varphi}{2})}{1+\sqrt{2}\cos\alpha\sin\alpha[\sin\frac{\varphi}{2}\sin(\frac{\varphi}{2}+\delta)-\cos\frac{\varphi}{2}\cos(\frac{\varphi}{2}-\delta)]}\\
\nonumber&\Bo{Fig.12 (C)}\\
\nonumber&P_Q(\ket{V}_s)=P_{S=\ket{V}|C=\ket{-},A=\ket{\alpha^{\bot}}}(\varphi,\alpha,\delta)\\
&=\frac{\frac{1}{2}\cos^{2}\alpha+\sin^{2}\alpha\cos^{2}\frac{\varphi}{2}-\sqrt{2}\cos\alpha\sin\alpha\cos\frac{\varphi}{2}\cos(\frac{\varphi}{2}-\delta)}{1+\sqrt{2}\cos\alpha\sin\alpha[\sin\frac{\varphi}{2}\sin(\frac{\varphi}{2}+\delta)-\cos\frac{\varphi}{2}\cos(\frac{\varphi}{2}-\delta)]},
\end{align}
where $\varphi$ is the relative phase between the two paths of MZI; $\alpha$ is the angle of projection angle; $\delta$ is the phase in $\rho_{CA}$ and equals to $0$ in Fig.11(\textbf{C}), 12(\textbf{C}).\\
$P_{S=\ket{H}|A=\ket{\alpha^{\bot}}}(\varphi,\alpha)$ ($P_{C}$ in main text) shows a classical mixture of $\ket{\textbf{w}}$ and $\ket{\textbf{p}}$; \\
$P_{S=\ket{H}|C=\ket{-},A=\ket{\alpha^{\bot}}}(\varphi,\alpha,\delta)$ ($P_{Q}$ in main text) shows a quantum superposition of  $\ket{\textbf{w}}$ and $\ket{\textbf{p}}$.\\
$P_{S=\ket{V}|A=\ket{\alpha^{\bot}}}(\varphi,\alpha)=1-P_{S=\ket{H}|A=\ket{\alpha^{\bot}}}(\varphi,\alpha)$;\\
$P_{S=\ket{V}|C=\ket{-},A=\ket{\alpha^{\bot}}}(\varphi,\alpha,\delta)$\\
$=1-P_{S=\ket{H}|C=\ket{A},A=\ket{\alpha^{\bot}}}(\varphi,\alpha,\delta)$.

We obtain the above probabilities from the experimental coincidence counts obtained in the experiment:
\begin{align}
\nonumber
\Bo{Fig.11 (B)}\ \ &P_{S=\ket{H}|A=\ket{\alpha^{\bot}}}(\varphi,\alpha)\\
&=\frac{C_{H+\alpha^{\bot}}+C_{H-\alpha^{\bot}}}{C_{H+\alpha^{\bot}}+C_{V+\alpha^{\bot}}+C_{H-\alpha^{\bot}}+C_{V-\alpha^{\bot}}} \\
\nonumber
\Bo{Fig.12 (B)}\ \ &P_{S=\ket{V}|A=\ket{\alpha^{\bot}}}(\varphi,\alpha)\\
&=\frac{C_{V+\alpha^{\bot}}+C_{V-\alpha^{\bot}}}{C_{H+\alpha^{\bot}}+C_{V+\alpha^{\bot}}+C_{H-\alpha^{\bot}}+C_{V-\alpha^{\bot}}} 
\end{align}
\begin{align}
\nonumber
\Bo{Fig.11 (D)}\ \     &P_{S=\ket{H}|C=\ket{-},A=\ket{\alpha^{\bot}}}(\varphi,\alpha,\delta)\\
&=\frac{C_{H-\alpha^{\bot}}}{C_{H-\alpha^{\bot}}+C_{V-\alpha^{\bot}}} \\
\nonumber
\Bo{Fig.12 (D)}\ \   &P_{S=\ket{V}|C=\ket{-},A=\ket{\alpha^{\bot}}}(\varphi,\alpha,\delta)\\
&=\frac{C_{V-\alpha^{\bot}}}{C_{H-\alpha^{\bot}}+C_{V-\alpha^{\bot}}}, 
\end{align}
where $C_{ijk}$ represent the three-fold coincidence counts of photons S,C,A with polarization i,j,k conditionally on the detection trigger photon T, respectively.\\
\begin{figure*}[h]
\includegraphics[width=12cm]{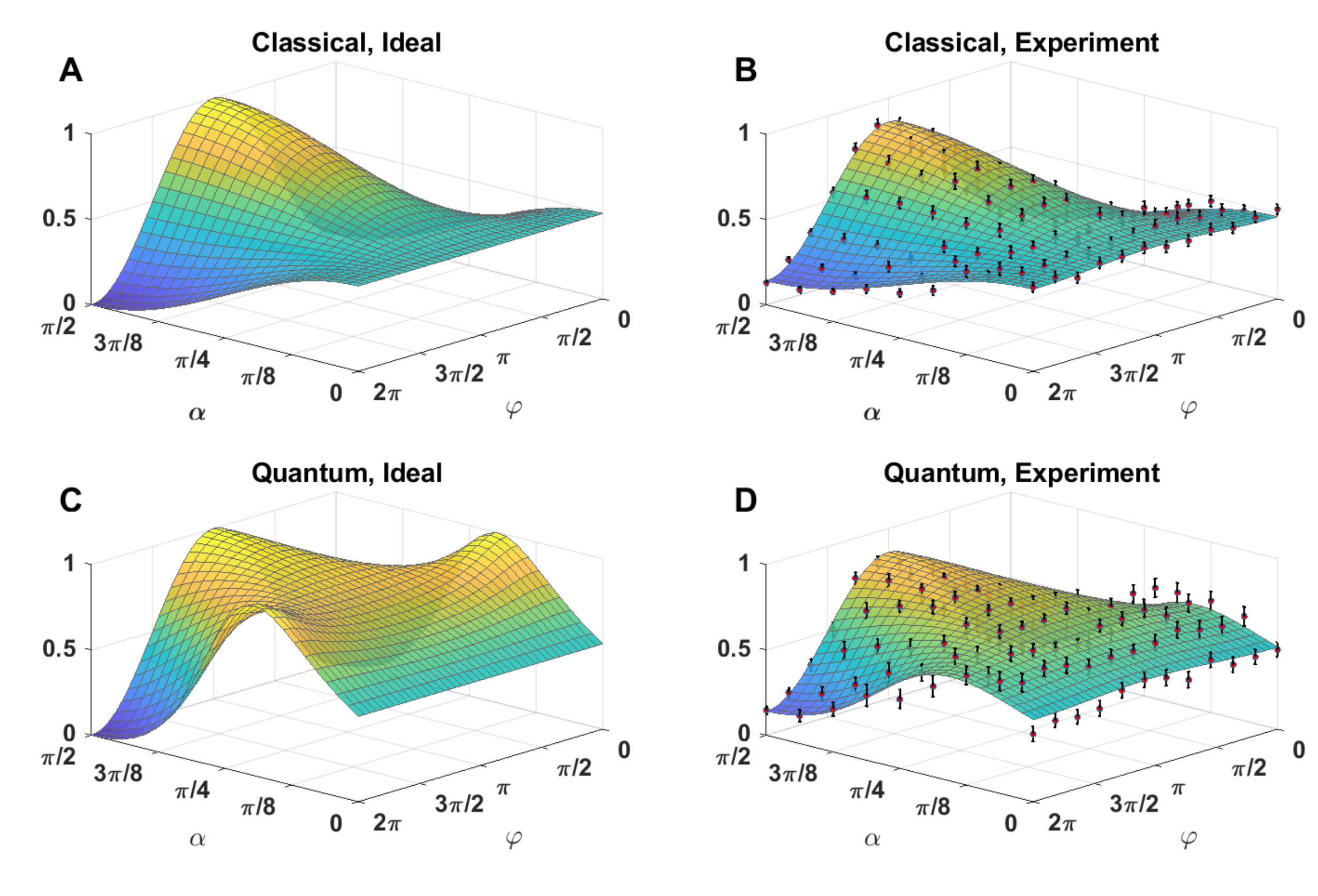}
\caption{Demonstration of continuous transitions between particle and wave states in both classical and quantum scenarios. (\textbf{A}) Simulated ideal and (\textbf{B}) measured probabilities $P_{S=\ket{H}|A=\ket{\alpha^{\bot}}}(\varphi,\alpha)$ for a classical mixture of particle and wave states. We scan the phase $\varphi$ in MZI for photon S and polarization rotation angle $\alpha$ of photon A. The experimental data are shown in red dots. (\textbf{C}) Simulated ideal and (\textbf{D}) measured probability $P_{S=\ket{H}|C=\ket{A},A=\ket{\alpha^{\bot}}}(\varphi,\alpha,\delta)$ for a quantum coherent superposition of particle and wave states. Note that the parameters used to calculate the surface plots for probability distributions shown in (\textbf{B}) and (\textbf{D}) are based on the values obtained from the independent characterizations of our photon-pair sources, CH gate and polarization contrast. The error bars are derived from Poissonian statistics and error propagations.}
\end{figure*}

\begin{figure*}[h]
\includegraphics[width=12cm]{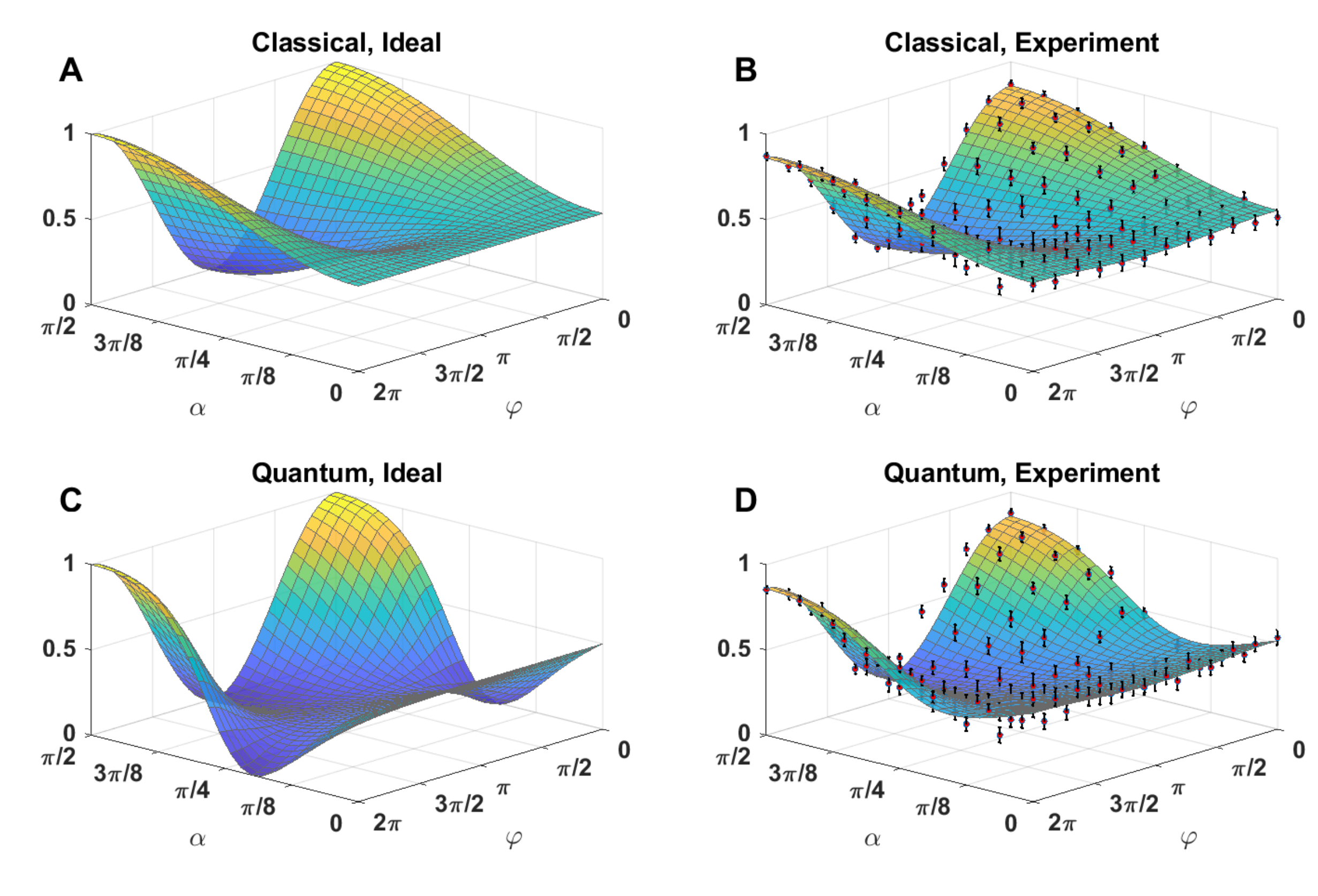}
\caption{Complementary results of Fig.11. (\textbf{A}) Simulated ideal and (\textbf{B}) measured probability $P_{S=\ket{V}|A=\ket{\alpha^{\bot}}}(\varphi,\alpha)$ for a classical mixture of particle and wave states, where the phase $\varphi$ in MZI for photon S and polarization rotation angle $\alpha$ of photon A are scanned. The experimental data (red dots) and the theoretical predictions (surface plots) show good agreement. (\textbf{C}) Simulated ideal and (\textbf{D}) measured probability $P_{S=\ket{V}|C=\ket{A},A=\ket{\alpha^{\bot}}}(\varphi,\alpha,\delta)$ for a quantum coherent superposition of particle and wave states. Note that the parameters used to calculate the surface plots for probability distributions shown in (\textbf{B}) and (\textbf{D}) are based on the values obtained from independent experimental measurements. The error bars are derived from Poissonian statistics and error propagations.}
\end{figure*}

As mentioned in the main text, the most direct proof for the quantum nature of the wave--particle superposition is to show that the result is sensitive to the relative phase between the wave and particle states, $\delta$. In experiment, we fix $\alpha=\frac{\pi}{4}$ and measure $P_C(\ket{H}_s)$, $P_Q(\ket{H}_s)$, $P_C(\ket{V}_s)$ and $P_Q(\ket{V}_s)$  to show their different dependence on phase $\delta$. When $\alpha=\frac{\pi}{4}$, Eq.D1-D4 give
\begin{align}
P_C(\ket{H}_s)&=P_{S=\ket{H}|A=\ket{\alpha^{\bot}}}(\varphi)=\frac{1}{4}+\frac{1}{2}\sin^{2}\frac{\varphi}{2}\\
P_C(\ket{V}_s)&=P_{S=\ket{V}|A=\ket{\alpha^{\bot}}}(\varphi)=\frac{1}{4}+\frac{1}{2}\cos^{2}\frac{\varphi}{2}\\
\nonumber
P_Q(\ket{H}_s)&=P_{S=\ket{H}|C=\ket{-},A=\ket{\alpha^{\bot}}}(\varphi,\delta)\\
&=\frac{\frac{1}{4}+\frac{1}{2}\sin^{2}\frac{\varphi}{2}+\frac{1}{\sqrt{2}}\sin\frac{\varphi}{2}\sin(\delta+\frac{\varphi}{2})}{1+\frac{1}{\sqrt{2}}[\sin\frac{\varphi}{2}\sin(\frac{\varphi}{2}+\delta)-\cos\frac{\varphi}{2}\cos(\frac{\varphi}{2}-\delta)]}\\
\nonumber
P_Q(\ket{H}_s)&=P_{S=\ket{H}|C=\ket{-},A=\ket{\alpha^{\bot}}}(\varphi,\delta)\\
&=\frac{\frac{1}{4}+\frac{1}{2}\cos^{2}\frac{\varphi}{2}-\frac{1}{\sqrt{2}}\cos\frac{\varphi}{2}\cos(\frac{\varphi}{2}-\delta)}{1+\frac{1}{\sqrt{2}}[\sin\frac{\varphi}{2}\sin(\frac{\varphi}{2}+\delta)-\cos\frac{\varphi}{2}\cos(\frac{\varphi}{2}-\delta)]}
\end{align}
The results are shown in Fig.13-14. Note that the data presented in Fig.13 (\textbf{B}) and (\textbf{D}) are identical to that shown in the Fig.4 (\textbf{C}) and (\textbf{F}) in the main text.

\begin{figure*}[!ph]
\includegraphics[width=12cm]{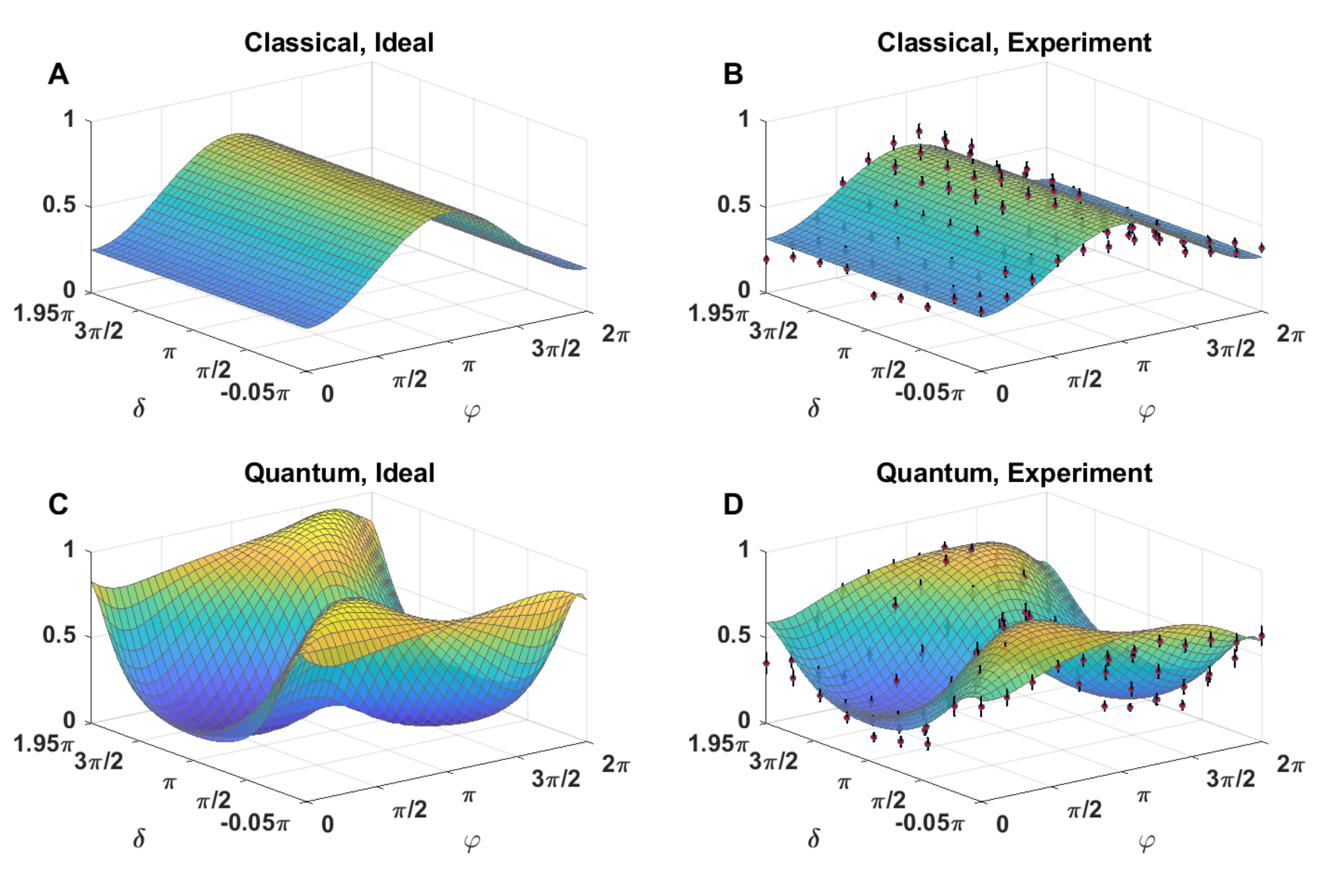}
\caption{Witnessing and controlling wave--particle quantum superpositions. We fix $\alpha=\frac{\pi}{4}$ and measure $P_{C}(\ket{H}_s)$ and $P_{Q}(\ket{H}_s)$ as functions of the phase $\varphi$ for photon S and the phase $\delta$ between the wave and the particle states of photon S, respectively.  (\textbf{A}) Simulated ideal and (\textbf{B}) measured probability $P_{C}(\ket{H}_s)$ for a classical mixture of particle and wave states. (\textbf{C}) Simulated ideal and (\textbf{D}) measured probability $P_{Q}(\ket{H}_s)$ for a quantum coherent superposition of particle and wave states. $P_{C}(\ket{H}_s)$ is clearly independent of $\delta$, whereas $P_{Q}(\ket{H}_s)$ is strongly dependent on $\delta$, manifesting the quantum nature of the superposition of wave and particle states. Note that the parameters used to calculate the surface plots for probability distributions shown in (\textbf{B}) and (\textbf{D}) are based on the values obtained from independent experimental measurements. The error bars are derived from Poissonian statistics and error propagations.}
\end{figure*}

\begin{figure*}[!h]
\includegraphics[width=12cm]{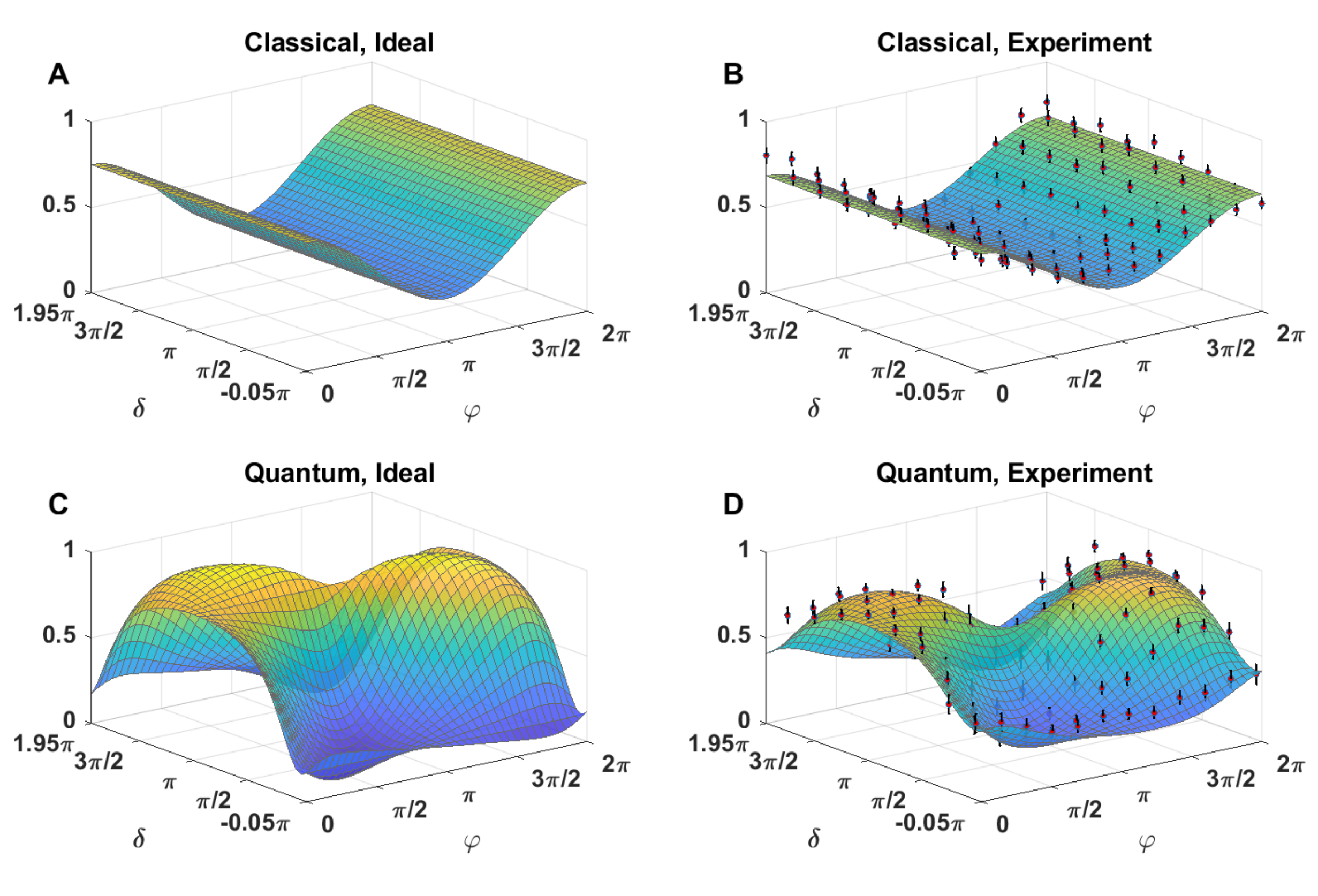}
\caption{Complementary results of Fig.13. (\textbf{A}) Simulated ideal and (\textbf{B}) measured probability $P_{C}(\ket{V}_s)$ for a classical mixture of particle and wave states. (\textbf{C}) Simulated ideal and (\textbf{D}) measured probability $P_{Q}(\ket{V}_s)$ for a quantum coherent superposition of particle and wave states. $P_{C}(\ket{V}_s)$ is clearly independent of $\delta$, whereas $P_{Q}(\ket{V}_s)$ is strongly dependent on $\delta$, manifesting the quantum nature of the superposition of wave and particle states. Note that the parameters used to calculate the surface plots for probability distributions shown in (\textbf{B}) and (\textbf{D}) are based on the values obtained from independent experimental measurements. The error bars are derived from Poissonian statistics and error propagations.}
\end{figure*}

\subsection{Data fitting}

One of the main contributions reducing the contrast of our results is the multiphoton emission from SPDC source. Here we introduce parameters to describe the imperfect state. The ideal state of the generated photon pairs are:
\begin{align}
&\rho_{ST}=\ket{VH}\bra{VH}_{ST}\\
&\rho_{CA}=\frac{1}{2}(\ket{HV}+e^{i\delta}\ket{VH})(\bra{HV}+e^{-i\delta}\bra{VH})_{CA}.
\end{align}
In our analysis, they are approximated to be Werner states~\cite{PhysRevA.40.4277}:
\begin{align}
&\rho'_{ST}=F_{1}\rho_{ST}+\frac{1-F_{1}}{4}\Bo{I}\\
&\rho'_{CA}=F_{2}\rho_{CA}+\frac{1-F_{2}}{4}\Bo{I}
\end{align}
with fidelities $F_{1},F_{2}$, which can be obtained via experimental results. Multiphoton emission reduces both fidelities. Before entering the CZ gate, photon S passes through Hadamard gate ($\Bo{H_{S}}$), SBC ($\Bo{Phi_{S}}$), W gate ($\Bo{W_{S}}$) sequentially. Photon A passes $\alpha$ gate ($\Bo{Alpha_{A}}$). We denote all the operations above as $\Bo{M}_{1}$ and the current state as $\rho_{1}$:
\begin{align}
\Bo{M}_{1}&=\Bo{W}_{S}\Bo{Phi}_{S}\Bo{H}_{S}\otimes\Bo{Alpha}_{A}\otimes \Bo{I}_{CT}\\
\rho_{1}&=\Bo{M}_{1}\rho'_{ST}\otimes\rho'_{CA}\Bo{M}_{1}^{\dagger}
\end{align}
where
\begin{align}
\rm{Hadmard\ gate}: 
\Bo{H}_{S}=&\begin{bmatrix}
&\cos\pi/4 &\sin\pi/4\\
&\sin\pi/4 &-\cos\pi/4\\
\end{bmatrix}\\
\rm{Phase\ gate}: 
\Bo{Phi}_{S}=&\begin{bmatrix}
& 1 & 0\\
& 0 & e^{i\varphi}\\
\end{bmatrix}\\
\rm{\alpha\ gate}: 
\Bo{Alpha}_{A}=&\begin{bmatrix}
&\cos\alpha &\sin\alpha\\
&\sin\alpha &-\cos\alpha\\
\end{bmatrix}.
\end{align}

After the CZ gate, we denote the current state as $\rho_{2}$. Multiphoton noise and the imperfect optical components (such as PPBS, wave plates and so on) reduce the interference contrast in Mach-Zehnder interferometer. We approximate the effect as white noise. Under this approximation, the effect contributes to each item of coincidence counts equally. Adding white noise item to $\rho_{2}$ gives $\rho_{2}'$:
\begin{align}
\rho_{2}&=\Bo{CZ}_{SC}\otimes \Bo{I_{AT}}\rho_{1}[\Bo{CZ}_{SC}\otimes \Bo{I_{AT}}]^{\dagger}\\
\rho_{2}'&=F_{3}\rho_{2}+\frac{1-F_{3}}{16}\Bo{I},
\end{align}
where $F_{3}$ is related to the HOM interference contrast and hence the fidelity of the CZ gate.

Photon S passes another W gate after CZ gate (Eq. A1) and gives the final state $\rho_{3}$:
\begin{equation}
\Bo{\rho}_{3}=(\Bo{W}_{S}\otimes I_{CAT})\rho_{2}'(\Bo{W}_{S}\otimes I_{CAT})^{\dagger}
\end{equation}
Then we can obtain the following results as functions of the three parameters $F_{1},F_{2}$ and $F_{3}$: 
\begin{align}
\nonumber P_{H+\alpha^{\bot}}=&\sand{H+V}{\rho_{3}}{H+V}\\
\nonumber =&\frac{1}{32}[2F_{1}F_{3}(1-\cos\varphi+F_{2}\cos\varphi\cos2\alpha)\\
\nonumber -&\sqrt{2}F_{2}F_{3}(1+F_{1})\sin2\alpha\cos\delta\\
+&2\sqrt{2}F_{1}F_{2}F_{3}\sin2\alpha\cos(\delta+\varphi)+2];\\
\nonumber P_{V+\alpha^{\bot}}=&\sand{V+V}{\rho_{3}}{V+V}\\
\nonumber =&\frac{1}{32}[2F_{1}F_{3}(1+\cos\varphi-F_{2}\cos\varphi\cos2\alpha)\\
\nonumber +&\sqrt{2}F_{2}F_{3}(1+F_{1})\sin2\alpha\cos\delta\\
+&2\sqrt{2}F_{1}F_{2}F_{3}\sin2\alpha\cos(\delta-\varphi)+2];\\
\nonumber P_{H-\alpha^{\bot}}=&\sand{H-V}{\rho_{3}}{H-V}\\
\nonumber =&\frac{1}{32}[2F_{1}F_{3}(1-\cos\varphi+F_{2}\cos\varphi\cos2\alpha)\\
\nonumber +&\sqrt{2}F_{2}F_{3}(1+F_{1})\sin2\alpha\cos\delta\\
-&2\sqrt{2}F_{1}F_{2}F_{3}\sin2\alpha\cos(\delta+\varphi)+2];\\
\nonumber P_{V-\alpha^{\bot}}=&\sand{V-V}{\rho_{3}}{V-V}\\
\nonumber =&\frac{1}{32}[2F_{1}F_{3}(1+\cos\varphi-F_{2}\cos\varphi\cos2\alpha)\\
\nonumber -&\sqrt{2}F_{2}F_{3}(1+F_{1})\sin2\alpha\cos\delta\\
-&2\sqrt{2}F_{1}F_{2}F_{3}\sin2\alpha\cos(\delta-\varphi)+2],
\end{align}
where $P_{ijk}$ stands for the probability of obtaining photons S,C,A in i,j,k polarization, conditionally on the detection of photon T. \\
Based on the above analysis, we plot the theoretical predictions with the parameters value of $F_{1}=0.98$, $F_{2}=0.90$, $F_{3}=0.61$, as shown in surface plots in Fig.11-14 \textbf{B}, \textbf{D}.

\subsection{Comparison between the quantum superposition and the classical mixture of wave and particle states in a delayed-choice experiment}
To show the difference between quantum superposition and classical mixture of $\ket{\textbf{w}}$ and $\ket{\textbf{p}}$, we compared average value and value at $\varphi=0,\delta=0$ of them: $P_{C}=P_{S=\ket{H}|A=\ket{\alpha^{\bot}}}$ and $P_{Q}=P_{S=\ket{H}|C=\ket{-},A=\ket{\alpha^{\bot}}}$ which are calculated from experimental data. The results are in Fig.15 which shows a clear difference between $P_{C}$ and $P_{Q}$.

\begin{figure}[h]
\includegraphics[width=8cm]{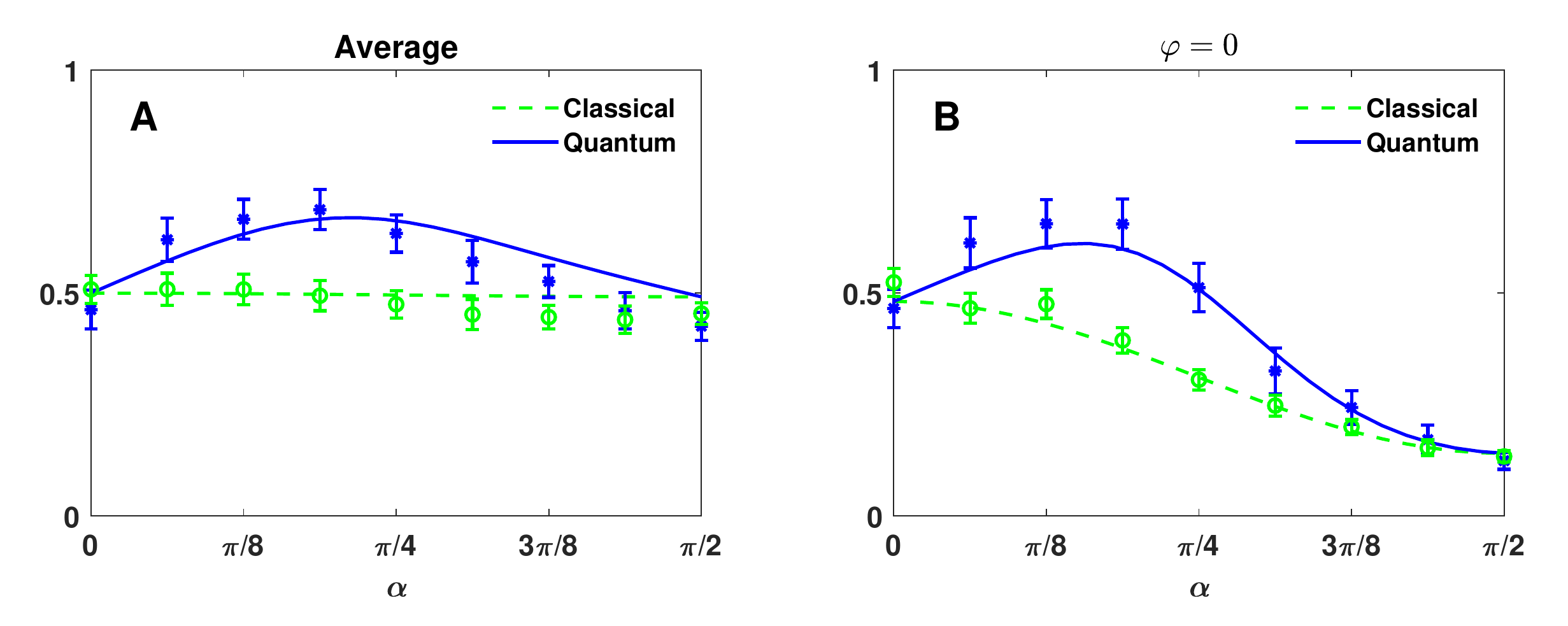}
\caption{\Bo{(A)} The arithmetic average value of experimental data ($P_{Q},P_{C}$) under the angle $\alpha$; \Bo{(B)} Value of $P_{Q},P_{C}$ at $\varphi=0, \delta=0$. The blue squares and green circles represent our experimental data for $P_{Q}$,$P_{C}$, respectively. The blue solid and green dash lines are the theoretical predictions.}
\end{figure}

\subsection{Experimental interference visibility of photon S without post-selecting the outcomes of photons C and A}
\begin{figure}
	\includegraphics[width=8cm]{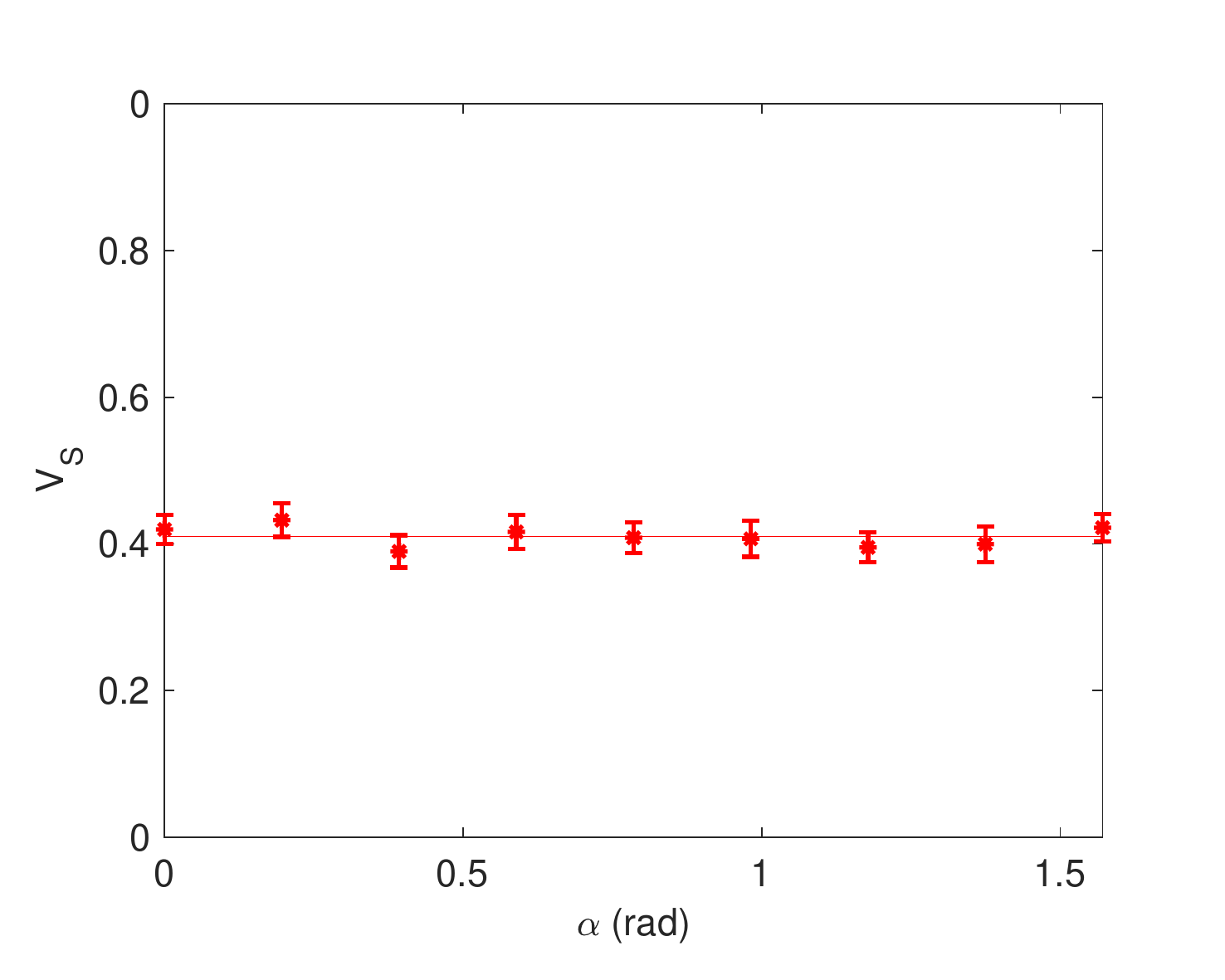}
	\caption{Visibility obtained from conditional probabilities of photon S without post-selecting on photons C and A by varying $\alpha$. Red line is the average value of visibilities. }
\end{figure}

In Ref \cite{ionicioiu2014wave}, it has been shown theoretically that one should obtain $\alpha$-independent interference visibilities if one ignore the outcomes of photons C and A. In our case, it is equivalent to trace out the polarization measurement results of both photons C and A. Ideally, we should obtain $V_{s}=0.5$. In Fig.12, we show our result with a nearly constant value 0.41. The discrepancy to the ideal value is mainly due to the limited interference visibility that we obtain experimentally.

\begin{equation}
\nonumber \ 
\nonumber \ 
\nonumber \ 
\nonumber \ 
\nonumber \ 
\nonumber \ 
\nonumber \ 
\nonumber \ 
\end{equation}



\nocite{*}
%

\end{document}